\documentclass{aastex631}

\usepackage{amsmath}
\usepackage{graphicx}
\usepackage{hyperref}
\usepackage{natbib}
\usepackage{CJK}
\usepackage[version=4]{mhchem}

\begin{document}
\begin{CJK*}{UTF8}{gbsn}

\title{Influence of Planetary Rotation on Supersonic Flow of Lava Planets: A Two-Dimensional Horizontal Model Analysis}

\correspondingauthor{Feng Ding}
\email{fengding@pku.edu.cn}

\author[0009-0001-9727-7908]{Zhuo-Yang Song}
\affiliation{School of Physics, Peking University, Beijing 100871, China}

\author[0000-0001-7758-4110]{Feng Ding}
\affiliation{Laboratory for Climate and Ocean-Atmosphere Studies, Department of Atmospheric and Oceanic Sciences, School of Physics, Peking University, Beijing 100871, China}

\author[0000-0002-4615-3702]{Wanying Kang}
\affiliation{Department of Earth, Atmospheric, and Planetary Sciences, Massachusetts Institute of Technology, 77 Massachusetts Ave., Cambridge, MA 02139, USA}




\begin{abstract}

The study of lava planets has attracted significant attention recently because of their close proximity to their host stars, which enhances their detectability for atmospheric characterization. Previous studies showed that the atmospheric flow becomes supersonic if the atmosphere was dominated by rocky vapor evaporated from the magma ocean around the substellar point of small lava planets. These studies often assumed an axisymmetric flow about the axis from the substellar point to the antistellar point but ignored the effect of planetary rotation on the climate. The spin rate of lava planets can be rather fast due to their close-in orbits, which can break the aforementioned symmetry and induce the asymmetric flow component. 
Here, we introduce a two-dimensional framework to explore the influence of planetary rotation on the atmospheric dynamics of these lava planets for the first time, and assess the sensitivity and range of application of our model. Starting from the established one-dimensional axisymmetric atmospheric solution, we obtain the governing equation for the asymmetric flow by expanding with respect to 1/Ro (Ro denotes Rossby number and exceeds unity for typical lava planets). The asymmetric component of supersonic flow is pivotal for future research on the observation of these atmospheres, flow patterns of the magma ocean currents driven by atmospheric winds, and deformation of the planetary shape over long timescales.

\end{abstract}



\section{Introduction}
More than 100 exoplanets have been observed to have masses and radii in the rocky-planet range and are so close to their host stars that the substellar point may be hot enough to melt the surface according to the NASA Exoplanets Archive \citep{Kite2016,Kang2023}. These hot rocky planets are tantalizing because they are much easier to detect and characterize than those in the habitable zone thanks to their short orbital period and high surface temperature \citep{Samuel2014observationCoRot7b,Demory2016,nguyen2020k2141b,Hu2024}. 

In general, atmospheric mass and composition on lava planets are controlled by the effective outgassing of gaseous species from the molten surface \citep{schaefer2012vaporization,wordsworth2018redox,Summary2022}, among which volatile components such as \ce{H2, He, H2O} may have been lost through atmospheric escape due to intense irradiation from the host star \citep{Valencia2010fateofCoRoT7b,kite2020secondary,Guo2024}.
Current observations suggest that the climate of lava planets may vary over a broad range. The thermal emission spectrum of lava planet 55~Cancri~e indicates that the planet still retains a thick gaseous envelope rich in \ce{CO2} or CO \citep{Hu2024}. On the other hand, the phase curve of lava planet K2-141b with symmetric shape and large amplitude \citep{K2} indicates a tenuous atmospheres that are dominated by more refractory mineral vapors (e.g., \ce{Na, SiO}). In this paper, we will focus on the second scenario, where the atmosphere is dominated by mineral vapor, and we aim to understand the asymmetric circulation component. 

Thus far, previous studies focusing on the mineral-vapor circulation on lava planets have adopted a one-dimensional (1D) axisymmetric framework \citep[e.g.,][]{Ingersoll1985,Castan2011,Kite2016,nguyen2020k2141b,Kang2021}. This model captures how mineral-vapor atmosphere may evaporate from the magma ocean on the hot dayside, forming a supersonic flow toward the nightside, forced by the extreme pressure gradient between dayside and nightside. 
This particular type of circulation regime is intrinsically identical to the sublimation-driven flow of the \ce{SO2} atmosphere on Io \citep{Ingersoll1985} and other local pure condensible atmospheres \citep{Ding2018}. In analogy to rocket nozzles and the solar wind \citep[e.g.,][]{landau1959fluidbook,parker1965solarwind}, the steady-state solution must pass through the sonic point smoothly, which was described as an atmospheric hydraulic jump in \citet{Ingersoll1985}. 
Tackling the singularity at the sonic point presents significant difficulties when attempting to numerically solve for the sublimation-/evaporation-driven flow. In these earlier studies, by assuming axis symmetry about the major axis ($z$-axis in Figure~\ref{fig:sketch}) and steady state, the governing equations are reduced to a set of 1D ordinary differential equations \citep[e.g.,][]{Ingersoll1985,Castan2011,Kang2021}, which allow semi-analytical solutions.

However, the assumption of an axisymmetric air flow pattern is only valid for slowly rotating planets when the influence of planetary rotation on atmospheric circulation is negligible \citep[see discussions in][]{merlis2010tidallylocked,showman2012circulation}, characterized by a large Rossby number. 
Lava planets are close to their host stars and exhibit short orbital and rotation periods due to tidal effects \citep{Castan2011,pierrehumbert2019tidelocked}. The Rossby number estimated in \cite{Castan2011} is only $\sim 2$ for typical lava planets, indicating that a notable asymmetric flow about the $z$-axis can be induced by the spin of the planet through the Coriolis effect.
The asymmetric component in the supersonic flow on lava planets is potentially important for understanding mass transport and deposition in mineral vapor atmospheres, long-term deformation of the planetary shape \citep{Kang2023}, and magma ocean circulation driven by surface wind stress and buoyancy contrast \citep{lai2024lavaocean}.

In this paper, we employ the perturbation method in the 1D axisymmetric model developed by \cite{Kang2021} and study the importance of the asymmetric component induced by the spin of lava planets in a two-dimensional (2D) framework for the first time. 
In Section~\ref{sec:theory}, we introduce our approach to perturb the 1D model and to solve the basic equations in a 2D framework. In Section~\ref{sec:results}, we present the asymmetric components of the surface air temperature, pressure, and wind velocity fields, and discuss how they vary with planetary parameters. In Section~\ref{sec:discussion}, we summarize our main conclusions.

\section{Two-dimensional Horizontal Model} \label{sec:theory}

\subsection{Simplification of the Problem}
We adopt the same simplifications as in previous studies, except for including the Coriolis force acting on horizontal flow. These general simplifications are briefly summarized as follows. 

\begin{figure}
\centering
\includegraphics[width=0.6\linewidth]{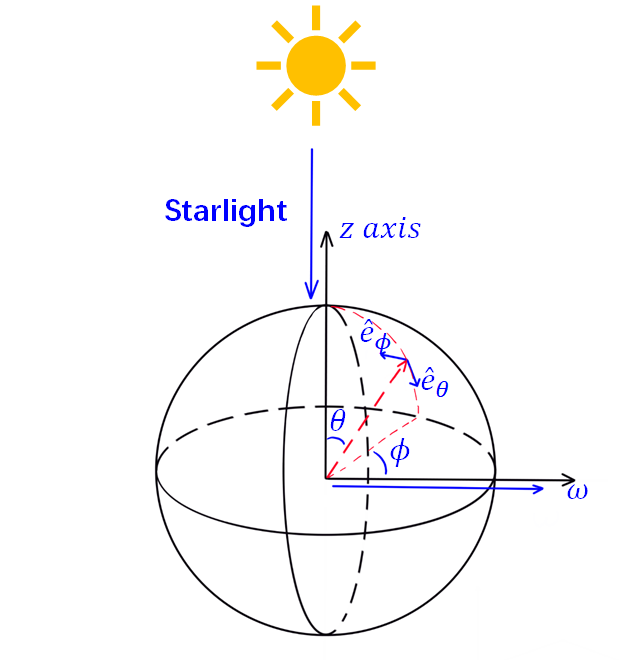}
\caption{\label{fig:sketch} Tidally locked coordinate ($\theta,\phi$) defined on lava planets in this work. The positive direction of transport flow $V$ and circulation flow $u$ are defined along the direction of $\theta$ and $\phi$, respectively. The angular velocity of planetary rotation points along $\theta=\pi/2$ and $\phi=0$. The host star is in the direction of $\theta=0$. We define the arc with $\phi = 0$ and $0<\theta<\pi/2$ as the prime meridian, then the hemisphere with $\sin{\phi}\leq0$ (i.e., $\pi\leq\phi<2\pi$) is the eastern hemisphere in our coordinate.}
\end{figure}

The lava planets are assumed to be in synchronous rotation. Based on the axis from the substellar point to the antistellar point, we define the spherical coordinate ($\theta,\phi$) used in this work in Figure~\ref{fig:sketch}, which is identical to the tidally locked coordinate defined in \citet{koll2015phasecurve}. Thus, the permanent day and night sides roughly correspond to the hemisphere with $\sin{\theta}>0$ and $\sin{\theta}<0$.

The mineral vapor atmosphere is assumed to be composed of a single chemical component and hydrostatically bound to the planet. The mineral vapor pressure in chemical equilibrium with the magma ocean $p_s$ is calculated using the formula $P_\text{chem}(T_s) = A_\text{chem} \exp({-B_\text{chem}/T_s})$ and the parameters $A_\text{chem}$ and $B_\text{chem}$ can be estimated by fitting the vapor pressure curve for the bulk silicate atmosphere model shown in Figure~1 of \citet{SodiumIsTheMost2009}. Only steady-state solutions are considered, and the time-varying transient motions are neglected.

The dominant energy balance on the surface of lava planets is established between the incoming radiative flux from the host star $F_\ast(\theta)$, the internal heat flux $F_i$, and the radiative cooling flux $\sigma T(\theta)^4$, where $\sigma$ is the Stefan-Boltzmann constant, because the energy exchange between the tenuous atmosphere and the surface is usually orders of magnitude smaller \citep{Kang2021}. The surface temperature can then be calculated as $T_{s}(\theta) = [(F_\ast(\theta) + F_i) / \sigma]^{1/4} $. Since lava planets are so close to their host stars, the incoming stellar radiation cannot be treated as parallel beams on other terrestrial planets, and more than half of the global area can be illuminated \citep{nguyen2020k2141b}. The angular distribution of $F_\ast(\theta)$ was discussed in detail and given in \citet{Kang2023}. In the complete dark region, which is less than half of the globe, the internal heat flux of the planet is assumed to be constant, and so is the surface temperature on the nightside $T_N = (F_i / \sigma)^{1/4} $. 

\subsection{Basic Equations}

Taking into account the Coriolis effect, the mass continuity, momentum and energy equation can be written as following in the steady state
\begin{eqnarray}
\nabla \cdot \left( \frac{\mathbf{V} P}{g} \right) &= {F}-D \label{eq:1},\\
\nabla \cdot \left( \frac{\mathbf{V} \mathbf{V} {P}}{g} \right) +  \nabla \left( \frac{\beta C_p T P}{g} \right) &= ({F}_- - D) \mathbf{V} - 2  {\omega} \sin \theta \cos \phi \left( \mathbf{n} \times \mathbf{V}   \right)  \frac{P}{g} \label{eq:2},\\
\nabla \cdot \left[ \left( \frac{|\mathbf{V}|^2}{2} + C_p T \right) \frac{\mathbf{V} {P}}{g} \right] &= D {L} + ({F}_- -D) \left( \frac{| \mathbf{V} |^2}{2} + C_p T \right) + {F}_+ C_p T_s, \label{eq:3}
\end{eqnarray}
where $\mathbf{n}$ represents the unit vector in the radial direction, $\mathbf{V} = V \mathbf{e}_\theta + u \mathbf{e}_\phi$ represents the horizontal velocity of the air flow ($V$ and $u$ are referred to as the transport flow and circulation flow hereafter), $P$ and $T$ represent the atmospheric pressure and temperature near the surface, $g$ represents the gravitational acceleration, $C_p$ represents the heat capacity at constant pressure per unit mass, $\omega$ represents the angular velocity of planetary rotation, $L$ represents the vaporization enthalpy per unit mass of the mineral vapor, $\beta = {R^\ast}/{(R^\ast + \mu C_p)}$ is a nondimensional constant where $R^\ast =8.31 \text{J mol}^{-1} \text{K}^{-1}$ is the universal gas constant and $\mu$ is the average molar mass for the atmosphere, $T_s$ represents the surface temperature. The gradient and divergence operators on the atmospheric fields are given by 

\begin{eqnarray}
    \nabla \cdot \mathbf{V} & = \frac{1}{a \sin{\theta}}\frac{\partial(V\sin{\theta})}{\partial \theta} + \frac{1}{a \sin{\theta}}\frac{\partial u}{\partial \phi},
    \label{eq:nabla1} \\
    {\nabla} & = \frac{\partial}{a\partial \theta}\mathbf{e}_\theta + \frac{1}{a\sin{\theta}}\frac{\partial}{\partial \phi} \mathbf{e}_\phi,
    \label{eq:nabla2}    
\end{eqnarray}
where $a$ is the planetary radius.

Various mechanisms enable the exchange of mass between the mineral vapor atmosphere and the surface, as indicated by the terms $D$ and $F$ \citep{Kang2021}. $D$ is the rate of mass condensation in the atmosphere when the near-surface air is supersaturated, and is calculated by raining out to keep the air pressure at the saturation vapor pressure $P_\text{sat}(T)$. \citet{Kang2021} demonstrated that the term $D$ plays a significant role in the energy conservation equation (Eq.~\ref{eq:3}), but it can be ignored in other governing equations.

$F$ is the flux of mass exchange determined by molecular collision and expressed as
\begin{eqnarray}
F(T_s, T, P) = \left\{
  \begin{array}{ll}
    \frac{\alpha P}{\sqrt{2\pi R^\ast T / \mu}} \left( \frac{P_{\text{chem}}(T_s)}{P} - 1 \right) & \text{within magma ocean, } T_s \geq T_m; \\
    \min \left\{ \frac{\alpha P}{\sqrt{2\pi R^\ast T / \mu}} \left( \frac{P_{\text{sat}}(T_s)}{P} - 1 \right),0 \right\} & \text{out of magma ocean, } T_s < T_m,
  \end{array}
\right.
\label{eq:F1}
\end{eqnarray}
where $T_m$ is the melting temperature of the rocky surface, $\alpha$ is the dimensionless sticking coefficient. Previous studies often assumed $\alpha=1$, indicating that each molecular collision results in molecule exchange to restore equilibrium. For a relatively warm surface, $\alpha$ could be less than unity due to kinetic effects (non-equilibrium fractionation) \citep{Hashimoto1990}. \citet{Kite2016} also pointed out that on lava planets, less refractory atmospheric constituents may form a diffusion barrier to surface condensation.
The mass fluxes can carry momentum and energy either from the atmosphere to the surface or in reverse. To make the basic equations clear, we further define $F_+, F_-$ as 
\begin{eqnarray}
\left\{
  \begin{array}{l}
    F_- = \min\{F,0\}; \\
    F_+ = \max\{F,0\},
  \end{array}
\right.
\label{eq:F2}
\end{eqnarray}
where $F_-$ and $F_+$ represent the flux of mass exchange from the atmosphere to the surface and from the surface to the atmosphere, respectively. 
Note that discontinuity in the horizontal distributions of $D$ and $F$ (Eqs.~\ref{eq:F1}-\ref{eq:F2}) will lead to nonsmoothness in atmospheric temperature and wind solutions, as shown in Figure~1 of \citet{Kang2023}.

The Coriolis force in the momentum equation (Eq.~\ref{eq:2}) will break the symmetry around the $z$-axis as assumed in previous studies. When ignoring the Coriolis force and integrating all equations along the tidally locked longitude $\phi$ in Figure~\ref{fig:sketch}, the basic equations are reduced to the conventional axisymmetric form \citep[e.g., Eqs.~B1-B3 in][]{Kang2021}.

\subsection{Nondimensionlization and Perturbation of Basic Equations}
Let the characteristic atmospheric pressure and temperature be $P_0$ and $T_0$, the physical variables in the basic equations can be nondimensionalized as
\begin{eqnarray}
\begin{array}{cc}
(\tilde{P},\tilde{P}_{chem},\tilde{P}_{sat})  =  \left(\frac{P}{P_0},\frac{P_{chem}}{P_0},\frac{P_{sat}}{P_0}\right), 
(\tilde{T}, \tilde{T_s}) =  \left( \frac{T}{T_0}, \frac{T_s}{T_0} \right), \\
\tilde{\bf{V}} = \frac{\bf{V}}{\sqrt{C_p T_0}},
\tilde{{\omega}} =  \frac{{\omega} a}{\sqrt{C_p T_0}},
\tilde{L} =  \frac{L}{C_p T_0},     (\tilde{D}, \tilde{F})  = (D, F) \frac{ag}{\sqrt{C_p T_0} P_0},
\end{array}
\label{eq:dim1}
\end{eqnarray}
where the tilde symbol denotes the nondimensionalized variables. Note that $\tilde{\omega}$ can be considered as the inverse of the Rossby number, estimated using the sound speed.
The atmospheric variables can be further expanded in the series form with the nondimensional planetary angular velocity $\tilde{\omega}$, provided $\tilde{\omega}$ is a small parameter:
\begin{eqnarray}
\begin{array}{lllll}
\tilde{V}(\theta, \phi) & = \tilde{V}^{(0)}(\theta)  +& \tilde{\omega} \tilde{V}^{(1)}(\theta, \phi)  +& \tilde{\omega}^2 \tilde{V}^{(2)}(\theta, \phi) + & \cdots, \\
\tilde{u}(\theta, \phi) & =   & \tilde{\omega} \tilde{u}^{(1)}(\theta, \phi)+ & \tilde{\omega}^2 \tilde{u}^{(2)}(\theta, \phi)+& \cdots, \\
\tilde{P}(\theta, \phi) & = \tilde{P}^{(0)}(\theta) + & \tilde{\omega} \tilde{P}^{(1)}(\theta, \phi)+& \tilde{\omega}^2 \tilde{P}^{(2)}(\theta, \phi)+& \cdots, \\
\tilde{T}(\theta, \phi) & = \tilde{T}^{(0)}(\theta) + &  \tilde{\omega} \tilde{T}^{(1)}(\theta, \phi)+& \tilde{\omega}^2 \tilde{T}^{(2)}(\theta, \phi)+& \cdots, 
\end{array}
\label{eq:dim2}
\end{eqnarray}
where the zeroth-order components $O(\tilde{\omega}^0)$ represent the axisymmetric state, and the higher order terms are induced by the Coriolis force neglected by previous works.

Here we solve the governing equation in nondimensional form because the nondimensional solutions can easily be generalized across a wide range of planetary parameters. For example, we will show later that the first-order nondimensional solutions depend only on the zeroth-order nondimensional states and are irrelevant to the planetary rotation rate. Therefore, the first-order nondimensional solution in one reference simulation is applicable to any scenario with any planetary rotation rate and the fixed nondimensional zeroth-order state, provided that the nondimensional planetary angular velocity $\tilde{\omega}$ remains small. 

Substituting Eqs.~\ref{eq:dim1}-\ref{eq:dim2} into Eqs.~\ref{eq:1}-\ref{eq:3}, the basic equations can be expressed as a series in the small parameter $\tilde{\omega}$. 
The zeroth-order components automatically satisfy the governing equations in previous axisymmetric models \citep[e.g., ][]{Kang2021}:
\begin{eqnarray}
\begin{array}{ll}
\frac{1}{\sin \theta} \frac{\text{d}}{\text{d}\theta} \left(\tilde{V}^{(0)} \tilde{P}^{(0)} \sin \theta \right) &= \tilde{F}^{(0)},\\
\frac{1}{\sin \theta} \frac{\text{d}}{\text{d}\theta} \left\{ \left[ \left(\tilde{V}^{(0)}\right) ^2 + \beta \tilde{T}^{(0)} \right] \tilde{P}^{(0)} \sin \theta \right\} &= \beta \tilde{T}^{(0)} \tilde{P}^{(0)} \cot \theta + \tilde{F}_-^{(0)} \tilde{V}^{(0)},\\
\frac{1}{\sin \theta} \frac{\text{d}}{\text{d}\theta} \left\{ \left[ {\left( \tilde{V}^{(0)} \right)^2}/{2} + \tilde{T}^{(0)} \right] \tilde{V}^{(0)} \tilde{P}^{(0)} \sin \theta \right\} &= \tilde{D}^{(0)} \tilde{L} + \tilde{F}_-^{(0)} \left[ {\left( \tilde{V}^{(0)} \right)^2}/{2} + \tilde{T}^{(0)} \right] + \tilde{F}_+^{(0)} \tilde{T}_s,\\
\frac{\text{d}}{\text{d}\phi} \left( \beta \tilde{T}^{(0)} \tilde{P}^{(0)} \right) &= 0. \label{eq:1d4}
\end{array}
\end{eqnarray}

Combining the axisymmetric components at $O(\tilde{\omega}^0)$ and the exact form of the Coriolis force $- 2  {\omega} \sin \theta \cos\phi  \left( \mathbf{n} \times \mathbf{V}   \right)$, the asymmetric  components at $O(\tilde{\omega}^1)$ can be decomposed into the product of a sinusoidal function of $\phi$ and another unknown function of $\theta$ as a prior:
\begin{eqnarray}
\begin{array}{ll}
\tilde{u}^{(1)}(\theta, \phi)  &= -\tilde{u}_1(\theta) \cos \phi, \\
\tilde{V}^{(1)}(\theta, \phi)  &= \tilde{V}_1(\theta)  \sin \phi,\\
\tilde{P}^{(1)}(\theta, \phi)  &= \tilde{P}_1(\theta)  \sin \phi, \\
\tilde{T}^{(1)}(\theta, \phi)  &= \tilde{T}_1(\theta)  \sin \phi.
\end{array} \label{eq:wo4}
\end{eqnarray}
Then the basic equations at $O(\tilde{\omega}^1)$ can be rearranged as four linearized ordinary differential equations (ODEs) with respect to $\theta$:
\begin{eqnarray}
\tilde{\Xi}_1 
\begin{pmatrix}
\tilde{u}_1 \\
\tilde{V}_1 \\
\tilde{P}_1 \\
\tilde{T}_1
\end{pmatrix}
+ \frac{\text{d}}{\text{d}\theta} \left[ \tilde{\Xi}_2 \begin{pmatrix}
\tilde{u}_1 \\
\tilde{V}_1 \\
\tilde{P}_1 \\
\tilde{T}_1
\end{pmatrix}
\right] &= 
\begin{pmatrix}
-2 \tilde{V}^{(0)} \tilde{P}^{(0)} \sin^2\theta \\
0 \\
0 \\
\tilde{\delta D} \tilde{L} \sin \theta
\end{pmatrix} \label{eq:2d1},
\end{eqnarray}
and the coefficient matrices are
\begin{eqnarray}
    & \tilde{\Xi}_1 = \begin{pmatrix}
  \tilde{F}_-^{(0)} \sin \theta & 0 & \beta \tilde{T}^{(0)} & \beta \tilde{P}^{(0)} \\
  \tilde{P}^{(0)} & 0 & -\tilde{F}_{,\tilde{P}} \sin\theta & -\tilde{F}_{,\tilde{T}} \sin\theta \\
  \tilde{V}^{(0)} \tilde{P}^{(0)} & -\tilde{F}_{-}^{(0)} \sin \theta & -\beta \tilde{T}^{(0)} \cos \theta -\tilde{F}_{-,\tilde{P}} \tilde{V}^{(0)} \sin \theta & -\beta \tilde{P}^{(0)} \cos \theta -\tilde{F}_{-,\tilde{T}} \tilde{V}^{(0)} \sin \theta \\
  [\frac{(\tilde{V}^{(0)})^2}{2} + \tilde{T}^{(0)}] \tilde{P}^{(0)} & -\tilde{F}_-^{(0)} \tilde{V}^{(0)} \sin \theta & 
  \begin{array}{c}
  \substack{
  [- \tilde{F}_{+,\tilde{P}} \tilde{T}_s - \\   \tilde{F}_{-,\tilde{P}} (\tilde{T}^{(0)} + \frac{(\tilde{V}^{(0)})^2}{2})]\sin \theta}
  \end{array} & 
  \begin{array}{c}
  \substack{[- \tilde{F}_-^{(0)} - \tilde{F}_{+,\tilde{T}} \tilde{T}_s  - \\ \tilde{F}_{-,\tilde{T}} (  \tilde{T}^{(0)} + \frac{(\tilde{V}^{(0)})^2}{2})] \sin \theta}
  \end{array}
\end{pmatrix}, 
  \label{eq:xi1} \\
  &  \tilde{\Xi}_2 =\begin{pmatrix}
  -\tilde{V}^{(0)} \tilde{P}^{(0)} & 0 & 0 & 0 \\
  0 & \tilde{P}^{(0)} & \tilde{V}^{(0)} & 0 \\
  0 & 2\tilde{V}^{(0)} \tilde{P}^{(0)} & (\tilde{V}^{(0)})^2 + \beta \tilde{T}^{(0)} & \beta \tilde{P}^{(0)} \\
  0 & (3(\tilde{V}^{(0)})^2/2 + \tilde{T}^{(0)}) \tilde{P}^{(0)} & ((\tilde{V}^{(0)})^2/2 + \tilde{T}^{(0)}) \tilde{V}^{(0)} & \tilde{V}^{(0)} \tilde{P}^{(0)}
\end{pmatrix} \sin\theta, \label{eq:xi2}
\end{eqnarray}
where we employ the notation $X_{,y}$ to represent the partial derivative $\frac{\partial X}{\partial y}$ for simplicity. The coefficient matrices (Eqs.~\ref{eq:xi1} and \ref{eq:xi2}) are expressed by the axisymmetric components at $O(\tilde{\omega}^0)$, and all elements are irrelevant to planetary rotation. 
According to \cite{Kang2021}, the condensation flux in super-saturated atmosphere $D$ only contributes significantly to the energy conservation equation (Eq.~\ref{eq:3}), but can be ignored in the mass and momentum conservation equations. Therefore, the nondimensional condensation flux $\tilde{D}=\tilde{D}^{(0)} + \tilde{\delta D}$ is only considered in the energy conservation equations in our 2D model (Eqs.~\ref{eq:1d4} and \ref{eq:2d1}). Appendix~\ref{Appendix:condensation} discusses the calculation of $\tilde{\delta D}$ at $O(\tilde{\omega}^1)$.

This paper focuses only on the effects of first-order solutions because of constraints on paper length. A brief discussion on the convergence and linear stability to small perturbations in the first-order solutions is given in Appendices~\ref{Appendix:system error} and \ref{Appendix:relaxation}.
In theory, once the first-order terms are established, the methods for solving higher-order terms follow similar approaches, which will not be covered in this study. 

\subsection{Numerical Algorithm to Solve Basic Equations}
We use the 1D axisymmetric model developed by \citet{Kang2021} to solve the governing equations at $O(\tilde{\omega}^0)$. Then the coefficient matrices (Eqs.~\ref{eq:xi1} and \ref{eq:xi2}) can be calculated by the axisymmetric components at $O(\tilde{\omega}^0)$. In order to solve for the first-order components in Eq.~\ref{eq:2d1}, we use the {LSODA} algorithm from SciPy, which can handle rigid and non-rigid ODE systems under the same framework \citep{petzold1983lsoda}. 

The major difficulty in solving Eq.~\ref{eq:2d1} is that the coefficient matrix $\Xi_2$ is not a full rank matrix at the substellar point ($\theta=0$), antistellar point ($\theta=\pi$) and the sonic point where $V^2 = \beta C_p T/(1-\beta)$. As noted in the Introduction, the steady-state solution must pass through the sonic point smoothly \citep{Ingersoll1985}. In our model, it is achieved by employing a binary search technique to find the appropriate boundary condition at the substellar point (see Appendix~\ref{Appendix:boundary condition} for details). 

\subsection{Model Setup}

\begin{table}[htbp]
\centering
\caption{Parameters used for the standard simulation.}
\label{tab:parameters} 
\begin{tabular}{llc}
\hline
\hline
Symbol & Name & Value/Form \\

\hline
\multicolumn{3}{l}{Parameters of host star} \\
\hline

$T_{\text{star}}$ & Effective temperature of host star & 5172 K \\
$M_{\text{star}}$ & Mass of host star & 0.905 $M_\odot$ \\
$R_{\text{star}}$ & Radius of host star & 0.943 $R_\odot$ \\

\hline
\multicolumn{3}{l}{Parameters of lava planet \citep{Kang2021}} \\
\hline

$T_{s0}$ & Substellar surface temperature & {3200} K \\
$T_N$ & Nightside surface temperature & 50 K \\
$\rho_p$  & Bulk density & $\rho_{\oplus}$ \\
$M_{p}$ & Planetary mass & {0.05} $M_{\oplus}$ \\
$T_{\text{period}}$ &  Period of rotation& 18 hours \\
$\alpha$ & Sticking coefficient for surface-atmosphere exchange & 1 \\
$T_m$ & Melting temperature of rocky surface & 1673 K \\

\hline
\multicolumn{3}{l}{Parameters of sodium atmosphere \citep{Castan2011, Kang2021}} \\
\hline

$C_p$ & Heat capacity at constant pressure & 903.3 J kg$^{-1}$ K$^{-1}$ \\
$\mu$ & Molar mass & 0.023 kg mol$^{-1}$ \\
L & Vaporization enthalpy & 96.96 kJ mol$^{-1}$ \\
$A_{\text{sat}}$ & Parameter for saturated vapor pressure & $10^{9.54}$ Pa \\
$B_{\text{sat}}$ & Parameter for saturated vapor pressure & 12070.4 K \\
$P_{\text{sat}} (T)$ & Saturation vapor pressure & $A_{\text{sat}} \exp(-{B_{\text{sat}}}/{T})$ \\
$A_{\text{chem}}$ & Parameter for chemical equilibrium pressure above magma ocean & $10^{9.6}$ Pa \\
$B_{\text{chem}}$ & Parameter for chemical equilibrium pressure above magma ocean & 38000 K \\
$P_{\text{chem}} (T)$ & Chemical equilibrium pressure above magma ocean & $A_{\text{chem}} \exp(-{B_{\text{chem}}}/{T})$ \\

\hline
\hline
\end{tabular}
\end{table}

Our default setup models a small synchronously rotating lava planet orbiting a K0 type star with planetary mass $M_p = 0.05 M_{\oplus}$, substellar surface temperature $T_{s0} = 3200$ K and rotation period of 18 hours. For this specific case, the characteristic values of the atmospheric temperature and pressure for nondimensionalization are $T_0 = T_{s0} = 3200$ K, $P_0 = P_\text{chem}(T_0) = 2.77\times 10^4$ Pa. The nondimensional planetary angular velocity for perturbing the basic equations is $\tilde{\omega} =\omega a /\sqrt{C_p T_0} =  0.13$. The magnitude of $\tilde{\omega}$ indicates notable asymmetric components in the supersonic airflow induced by planetary rotation, and validates the perturbative technique involving the small parameter $\tilde{\omega}$ as a suitable method for identifying these asymmetric components. Other parameters used in the standard simulation are listed in Table~\ref{tab:parameters}. Section~\ref{subsec:global} presents the results of the standard simulation and Section~\ref{subsec:sensiti} discusses how the first-order state varies with external parameters such as planetary mass, density, sticking coefficient for surface-atmosphere exchange, and substellar surface temperature.

\section{Results} \label{sec:results}

\subsection{Global features of the standard simulation} \label{subsec:global}

\begin{figure}[htbp]
  \centering
  

  \includegraphics[width=\textwidth]{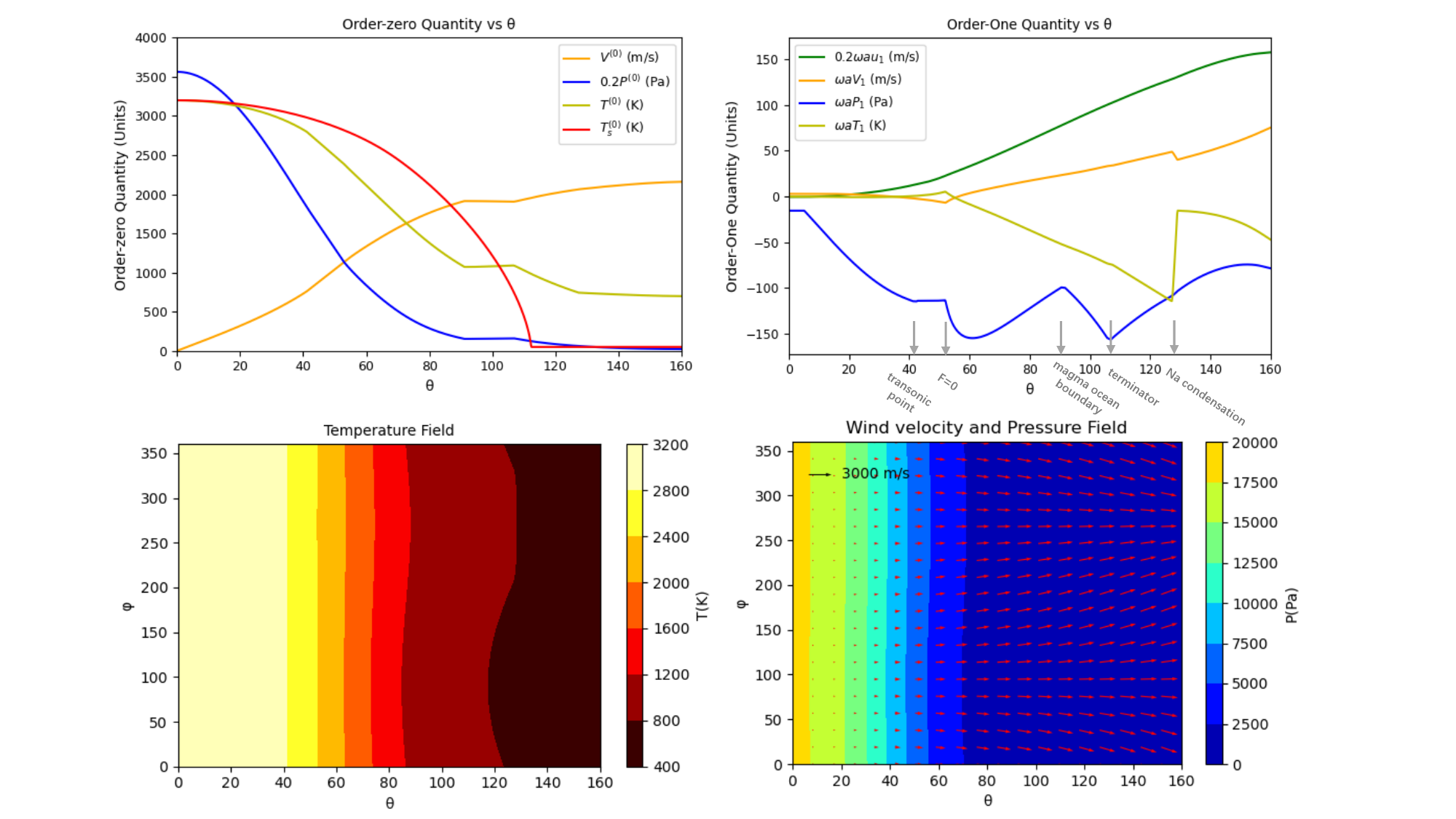}
  \caption{Upper panels: atmosphere fields as a function of tidally locked latitude $\theta$ for the zeroth-order state (left) and the first-order state (right) in the standard simulation, with dimension reinstated. Yellow curves represent the transport velocity $V$ along $\bf{e_\theta}$;  blue curves represent the surface pressure $P$; light green curves represent the atmospheric temperature $T$; red curve represents the surface temperature $T_s$ (left); dark green curve represents the circulation velocity $u$ along $\bf{e_\phi}$ (right). The physical mechanisms responsible for the non-smooth feature of the first-order solutions at different locations are provided below the horizontal axis. 
  Lower panels: surface distribution of $T_s$ (left), $P$ (contours in the right panel) and wind velocity (red vectors in the right panel). 
  Parameters used in the standard simulation are listed in Table~\ref{tab:parameters}.}
  \label{fig:show}
\end{figure}

Figure~\ref{fig:show} shows the zeroth- and first-order atmospheric state in the standard simulation, with dimension reinstated:
\begin{eqnarray}
\begin{array}{ll}
    u(\theta, \phi) &= - {\omega a} {u_1}(\theta) \cos \phi, \\
    V(\theta, \phi) &= V^{(0)}(\theta) + {\omega a} {V_1} (\theta) \sin \phi,\\
    P(\theta, \phi) &= P^{(0)}(\theta) + {\omega a} {P_1} (\theta) \sin \phi,\\
    T(\theta, \phi) &= T^{(0)}(\theta) + {\omega a} {T_1} (\theta) \sin \phi,
    \label{eq:define4}
\end{array}
\end{eqnarray}
where $V^{(0)},P^{(0)},T^{(0)}$ are zeroth-order solutions with dimension reinstated, and ${\omega a} u_1,{\omega a} V_1,{\omega a} P_1,{\omega a} T_1$ are first-order solutions with dimension reinstated \footnote{Note that $u_1, V_1$ are still dimensionless here, and the units for $P_1, T_1$ are Pa~m$^{-1}$~s and K~m$^{-1}$~s, respectively.}. The zeroth order components are the same as those produced by 1D axisymmetric models \citep[e.g.,][]{Kang2021}.
The surface pressure $P^{(0)}$ is almost always in chemical equilibrium with the magma ocean. The transport velocity $V^{(0)}$ increases from zero to over $10^3$~m~s$^{-1}$ driven by the strong pressure gradient between dayside and nightside, as internal energy of the fluid is converted into kinetic energy. The latent heat release associated with condensation provides additional energy for acceleration and slows the cooling rate. For more detailed discussions of the zeroth-order solution, we refer readers to \cite{Kang2021}. Here, we will focus on the first-order components.

The Coriolis effect in the momentum equation generates an azimuthal velocity $u$, by deflecting $V^{(0)}$. As can be seen in Eq.~\ref{eq:2}, the first-order term $- {\omega a} {u_1}(\theta) \cos \phi \mathbf{e_\phi}$ initially arises from the first-order Coriolis force $-2\omega \sin \theta \cos \phi V^{(0)}\mathbf{e_\phi}$, which leads to a monotonic increase of $u_1$ with $\theta$, as shown in the upper right panel of Figure~\ref{fig:show}. 
Then, the interactions between circulation and transport velocities will induce their higher-order components, as well as the asymmetric components of atmospheric pressure and temperature, through the mass and energy conservation equations. As a result, the substellar point symmetry assumed in previous studies is roughly applied for the region with $\theta < 60^\circ$ where the transport velocity $V$ is not strong enough. However, the atmospheric flow exhibits notable convergence westward and divergence eastward of the antistellar point. 

Generally speaking, the general circulation pattern that accounts for the first-order correction is similar to the atmospheric flow in the upper troposphere of tidally locked planets with the zero radiative timescale and a moderate drag timescale discussed in Appendix D of \citet{showman2011superrotation}, because the Coriolis effect plays a similar role in both cases. The shallow-water model developed by \citet{showman2011superrotation}, which assumes an incompressible atmosphere and excludes the interaction between the atmosphere and the surface, lacks the complexity required for application to lava planets characterized by mineral-vapor atmospheres. Nevertheless, their shallow-water model provided a distinct perspective on how planetary rotation breaks the substellar point symmetry on tidally locked planets. 

Another characteristic of the first-order components is the piecewise structure of their distributions along the tidally locked latitude $\theta$ (upper right panel of Figure~\ref{fig:show}). It stems from the piecewise nature of external forcings on the atmosphere (e.g., the non-smoothness of the stellar flux near the terminator $\theta \sim 110^\circ$, of the mass exchange flux near the magma ocean boundary $\theta \sim 90^\circ$, and of the sodium condensation flux in the atmosphere near $\theta \sim 130^\circ$) rather than our numerical solving algorithm. In our model, the mass exchange flux $F$ between the atmosphere and the surface is not continuous at the magma ocean boundary (Eqs.~\ref{eq:F1}-\ref{eq:F2}), which is responsible for the non-smooth feature of $P^{(0)}, T^{(0)},V^{(0)},P_1$. The transformation of a magma ocean into a rocky crust is not an instantaneous process \citep{herath2024thermallava}, and this gradual crystallization is not implemented in our model.


\subsection{Approximation of the circulation velocity $u_1$} 

\begin{figure}
\centering
\includegraphics[width=0.8\linewidth]{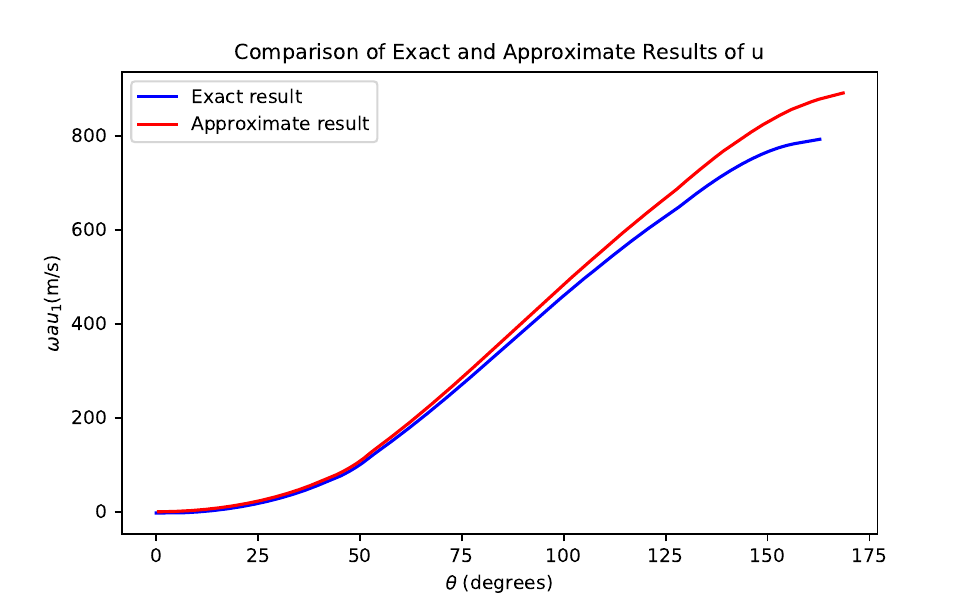}
\caption{ Comparison of the first-order circulation velocity $\omega a u_1(\theta)$ in the standard simulation between the exact solution (blue) and the approximation solution (red). The exact solution is the same as the green curve in the upper right panel of Figure~\ref{fig:show}, except for the rescaled vertical axis. }
\label{fig:approx}
\end{figure}

To understand the characteristics of $u_1$, especially its notable smoothness in contrast to the other three first-order components, we present an approximation solution. We start by identifying the following dominant balance of the first-order momentum equation:
\begin{eqnarray}
    \frac{\text{d} u_1}{ \text{d} \theta} & \approx \left\{
    \begin{array}{ll}
        -\frac{ F^{(0)} ag}{P^{(0)}V^{(0)}}u_1 + 2\sin \theta, & \text{where}\ F^{(0)}\geq 0;  \\
        2\sin \theta, & \text{where}\ F^{(0)}< 0.
    \end{array} 
    \right. 
    \label{eq:approx}
\end{eqnarray}
where the influence of first-order pressure and temperature anomalies, $P_1, T_1$, are neglected. As can be seen in Figure~\ref{fig:approx}, the above approximate solution $u_1$ generally matches the exact solution of our simulation except near the antistellar point.

Eq.~\ref{eq:approx} illustrates that the increase in $u_1$ with respect to the tidally locked latitude $\theta$ is driven by the Coriolis term $2\sin \theta$. In regions where the magma ocean supplies mineral vapor to the atmosphere ($F>0$), evaporation that carries zero momentum from the magma ocean acts as an equivalent drag that reduces $u_1$. Furthermore, the right-hand side of Eq.~\ref{eq:approx} is a continuous function with the tidally locked latitude $\theta$, which is responsible for the smooth behavior of $u_1$ in the upper right panel of Figure~\ref{fig:show}. 

\subsection{Application to lava planets with various parameters} \label{subsec:sensiti}

In this section, we further investigate how atmospheric circulation and temperature distributions (to be more exact, $u_1, V_1, P_1, T_1$ defined in Eq.~\ref{eq:define4}) vary with planetary parameters. The parameters we investigate include the substellar point temperature $T_{s0}$, the planetary mass $M_p$, the bulk density $\rho_p$, and the sticking coefficient for the surface-atmosphere exchange $\alpha$.

The nondimensional governing equations (Eqs.~\ref{eq:dim1}-\ref{eq:xi2}) show that $\alpha, M_p, \rho_p$ only influence the zeroth- and first-order solutions via the mass exchange flux term between the surface and atmosphere, taking the following simple form  
\begin{eqnarray}
    \alpha a g \propto \frac{\alpha}{1} \left(\frac{M_p}{M_{\oplus}}\right)^{\frac{2}{3}}\left(\frac{\rho_p}{\rho_{\oplus}}\right)^{\frac{1}{3}} = \left(\frac{M^*}{M_{\oplus}}\right)^{\frac{2}{3}}.
    \label{eq:packed}
\end{eqnarray}
Here we introduce a modified planetary mass $M^\ast\equiv \alpha^{3/2} (\rho_p/\rho_\oplus)^{1/2} M_p$ that accounts for the influence of $\alpha$ and $\rho_p$. In the rest of this section, we will see how $M^\ast$ will influence the solution. 

\begin{figure}[htbp]
  \centering
  \begin{minipage}{0.49\textwidth}
    \includegraphics[width=\textwidth]{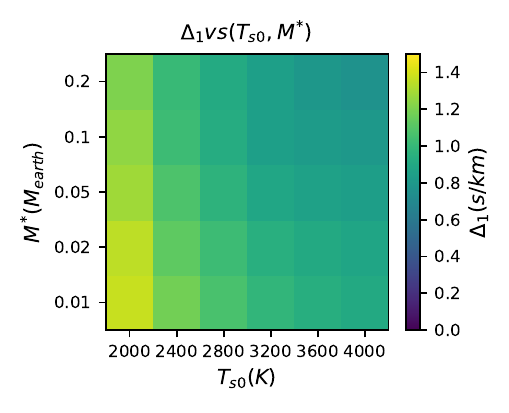}
  \end{minipage}%
  \hfill
  \begin{minipage}{0.49\textwidth}
    \includegraphics[width=\textwidth]{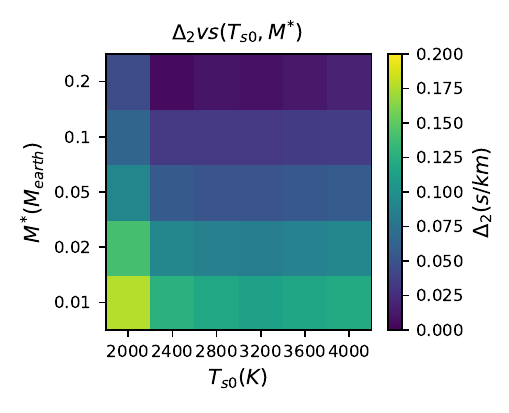}
  \end{minipage}
  
  \medskip 

  \begin{minipage}{0.49\textwidth}
    \includegraphics[width=\textwidth]{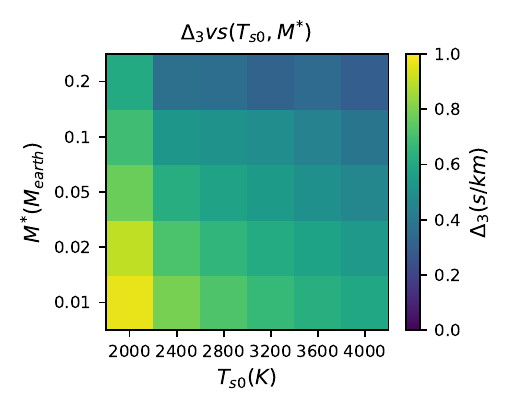}
  \end{minipage}%
  \hfill
  \begin{minipage}{0.49\textwidth}
    \includegraphics[width=\textwidth]{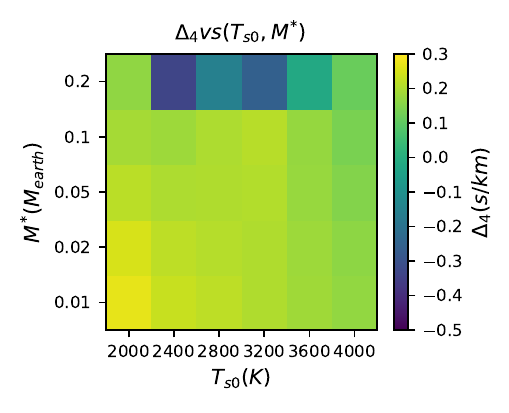}
  \end{minipage}



  \caption{Distribution of $\Delta_1,\Delta_2,\Delta_3,\Delta_4$ at $\theta = \pi/2$ as a function of $M^\ast$ and $T_{s0}$, when $M^\ast$ varies from $0.01M_\oplus$ and $0.2M_\oplus$ and the substellar temperature $T_{s0}$ varies from 2000 K to 4000 K.
  The outliers of $\Delta_4$ when $M^\ast = 0.2M_{\oplus}, T_{s0}$  = 2400 K, 2800 K, 3200 K, 3600 K are caused by the atmospheric condensation and associated latent heat release at $\theta=\pi/2$. }
  \label{fig:delta}
\end{figure}

To evaluate the relative magnitude of the first-order components with respect to the background axisymmetric states, we define four parameters as
\begin{eqnarray} \label{eq:delta}
    \Delta_1(\theta) = u_1 / V^{(0)}, \quad
    \Delta_2(\theta) = V_1 / V^{(0)}, \quad
    \Delta_3(\theta) = - P_1 / P^{(0)}, \quad
    \Delta_4(\theta) = - T_1 / T^{(0)}. \quad
\end{eqnarray}
Figure~\ref{fig:delta} illustrates the distribution of these four parameters at $\theta = \pi/2$ as a function of $M^\ast$ and $T_{s0}$, where $M^\ast$ varies from $0.01M_\oplus$ and $0.2M_\oplus$ and the substellar temperature $T_{s0}$ varies from 2000 K to 4000 K. As can be seen, the first-order corrections tend to be more significant for lava planets that are relatively small and cool, due to their greater $\tilde{\omega}$. A similar effect can be achieved with a weaker sticking coefficient $\alpha$. Specifically, the dependence of $\Delta_1(\pi/2) = u_1 / V^{(0)}$ on $M^\ast$ and $T_{s0}$ can be qualitatively explained by the ratio of the approximation form of $u_1(\theta)$ given by Eq.~\ref{eq:approx} and $V^{(0)} \lesssim \sqrt{C_p T_{s0}}$. As discussed previously, other first-order components are induced by $u_1$ and are therefore related to the strength of $u_1$, which accounts for the analogous behaviors of the four relative magnitudes $\Delta_i$ as a function of $M^\ast$ and $T_{s0}$.

\begin{figure}[htbp]
  \centering
  \begin{minipage}{0.3\textwidth}
    \includegraphics[width=\textwidth]{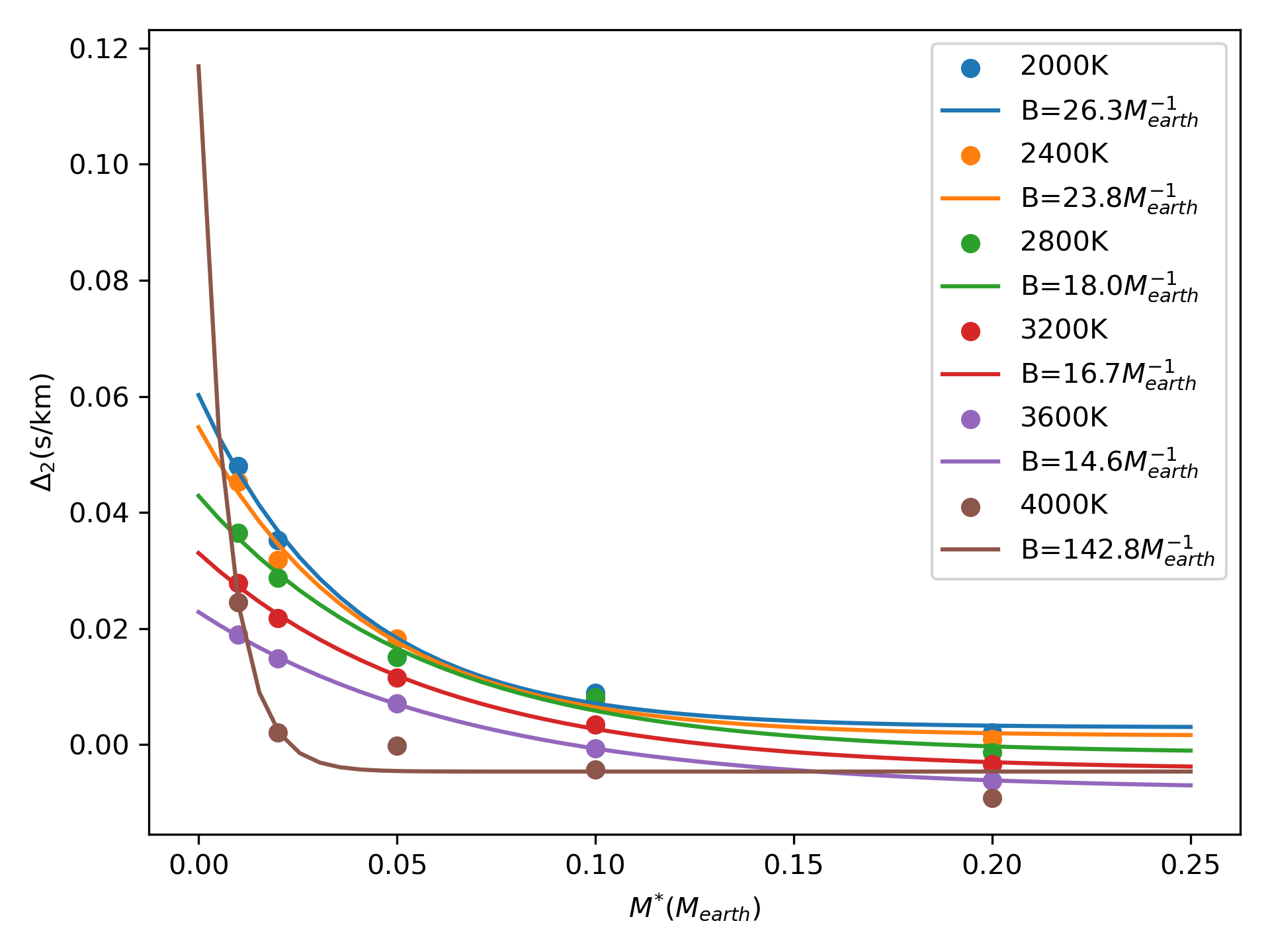}
  \end{minipage}%
  \hfill
  \begin{minipage}{0.3\textwidth}
    \includegraphics[width=\textwidth]{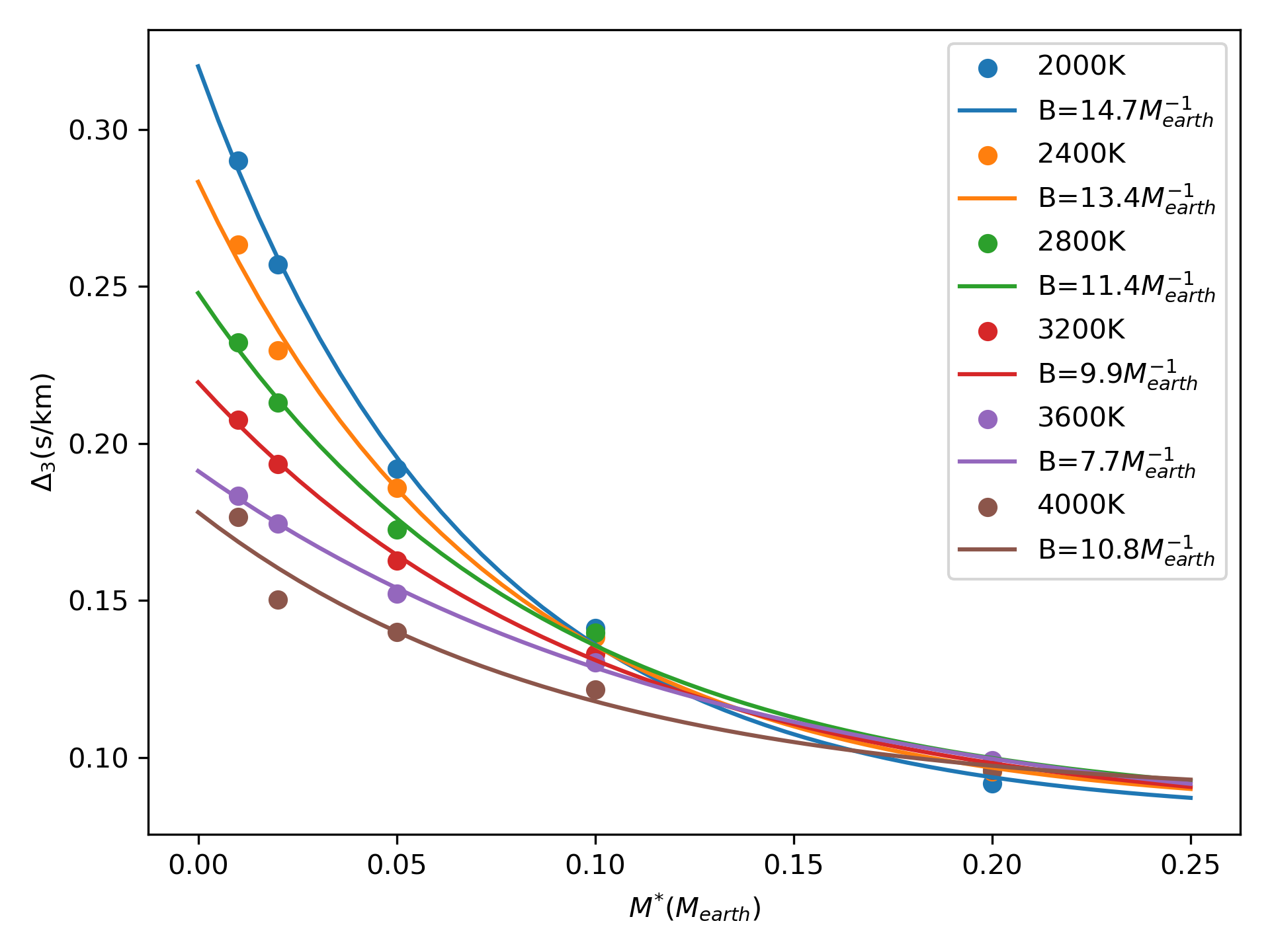}
  \end{minipage}
  \hfill
  \begin{minipage}{0.3\textwidth}
    \includegraphics[width=\textwidth]{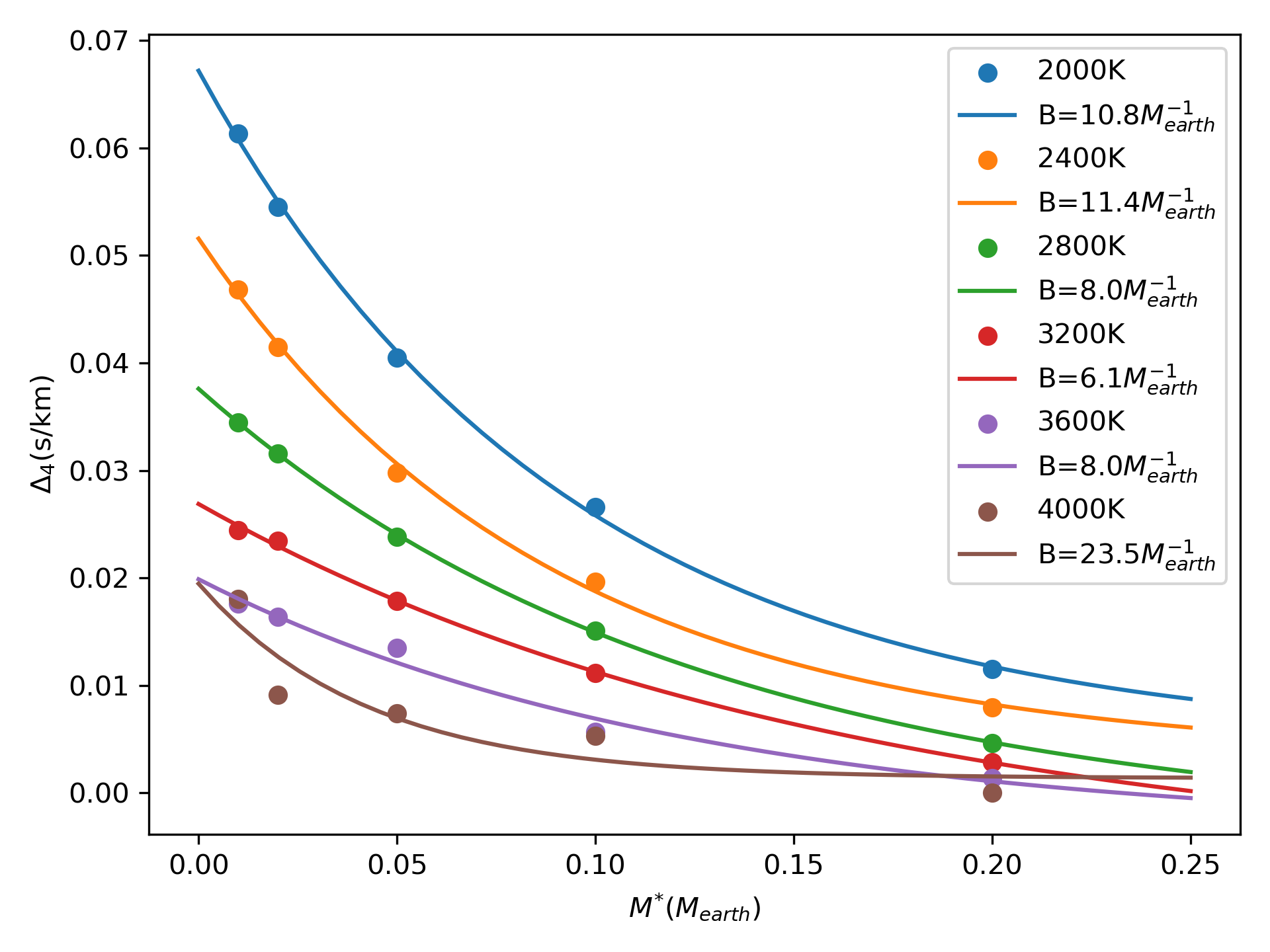}
  \end{minipage}
  
  \medskip 

  \begin{minipage}{0.3\textwidth}
    \includegraphics[width=\textwidth]{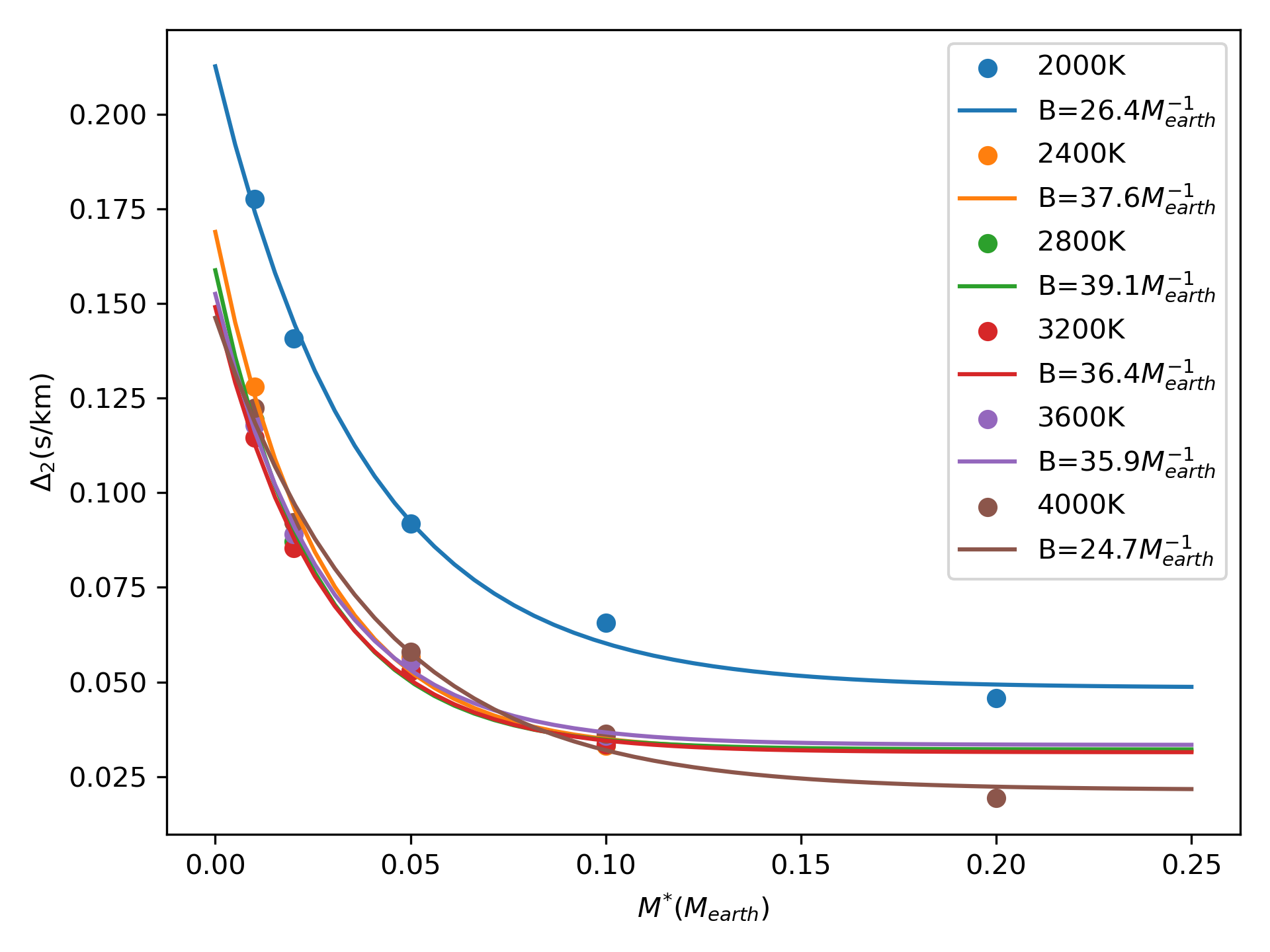}
  \end{minipage}%
  \hfill
  \begin{minipage}{0.3\textwidth}
    \includegraphics[width=\textwidth]{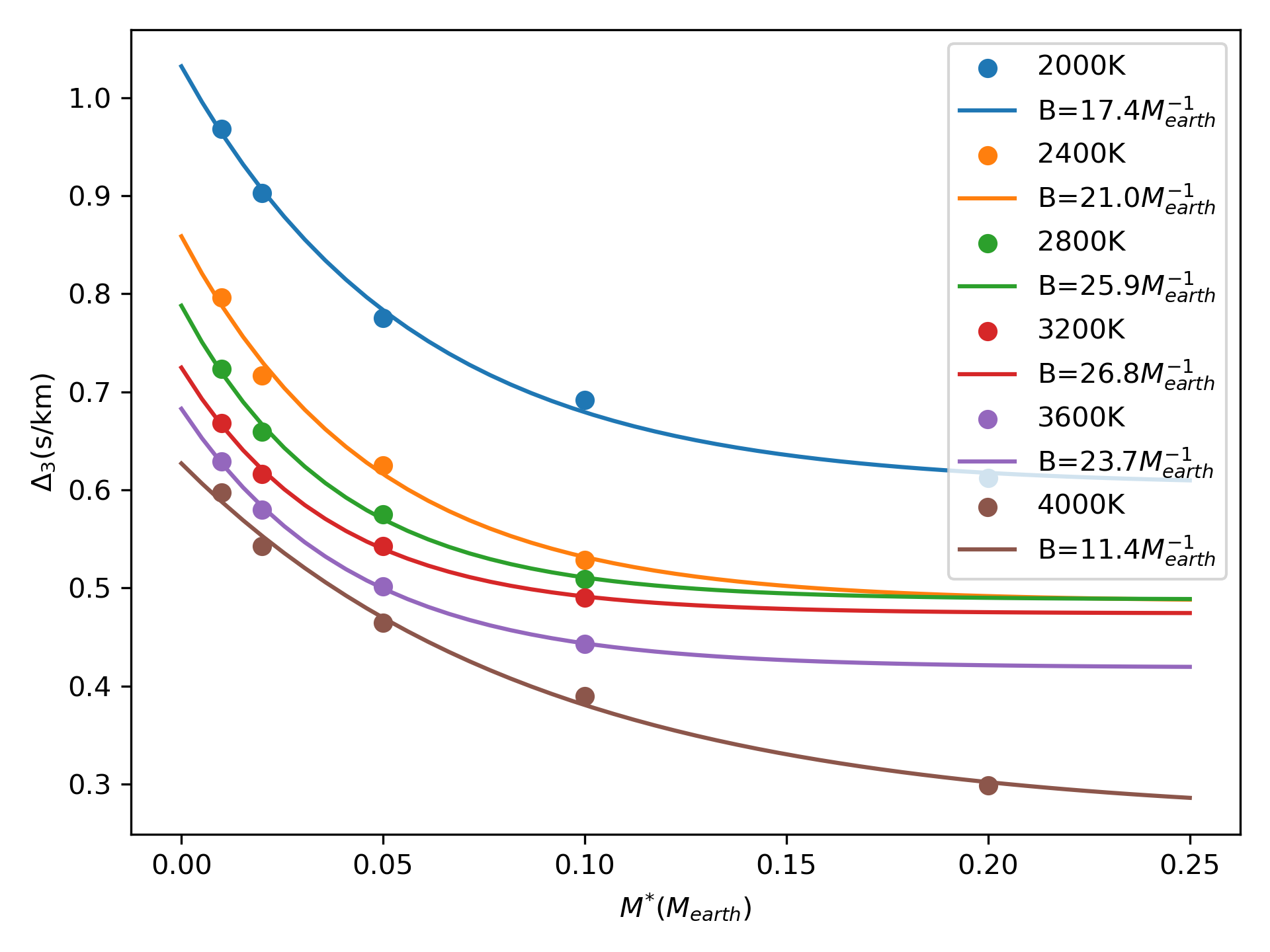}
  \end{minipage}
  \hfill
  \begin{minipage}{0.3\textwidth}
    \includegraphics[width=\textwidth]{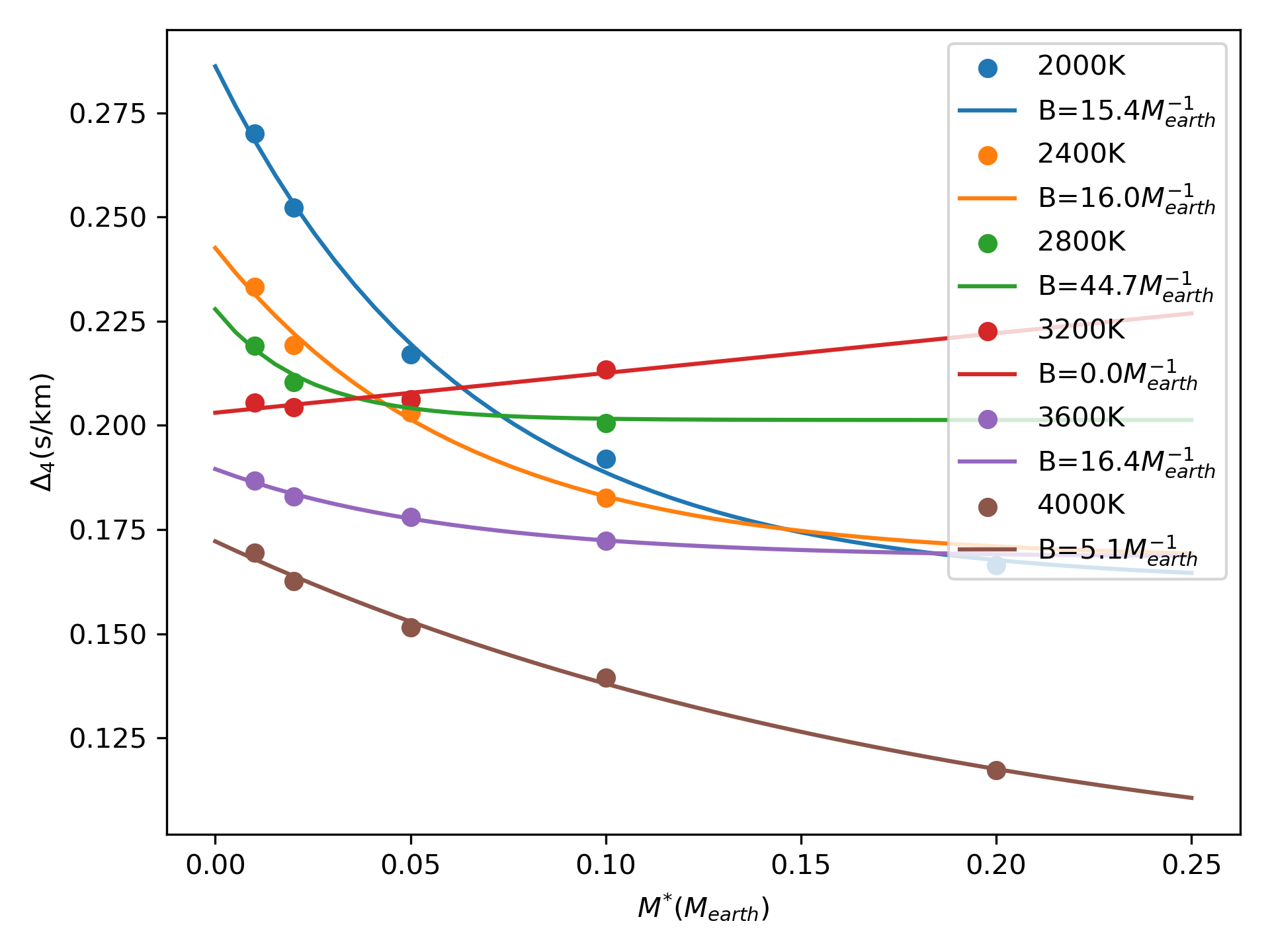}
  \end{minipage}

  \caption{$\Delta_2(\theta) = V_1 / V^{(0)}, 
    \Delta_3(\theta) = - P_1 / P^{(0)},
    \Delta_4(\theta) = - T_1 / T^{(0)}$ as a function of $M^\ast$ when $\theta = \pi/3$ (upper panels) and $\theta = \pi/2$ (lower panels). The numerical solutions are represented by colored dots with various substellar surface temperatures (blue, orange, green, red, purple, and brown points represent 2000 K,2400 K,2800 K,3200 K,3600 K, and 4000 K), and the best fits corresponding to each substellar surface temperature are represented by solid lines.  
    The fitted curves take the form of $\Delta_i = A_i \exp(-B_i M^\ast)+C_i, i=2,3,4$. The saturated points in the lower right panel of Figure~\ref{fig:delta} have been removed in these panels to avoid complication. }
\label{fig:stablized} 
\end{figure}

Due to the limitations of machine precision, our model is unable to obtain first-order solutions for lava planets more massive than Earth, such as K2-141b. However, the amplitudes $\Delta_i$ defined in Eq.~\ref{eq:delta} in our simulations can be approximately described by fitted curves in the form of $\Delta_i = A_i \exp(-B_i M^\ast)+C_i, i=2,3,4$ when $M^\ast/M_\oplus \leq 0.25$, as shown in Figure~\ref{fig:stablized}. We expect that first-order solutions on massive lava planets could be estimated by extrapolating the empirical form identified in Figure~\ref{fig:stablized} as long as $\tilde{\omega}$ remains small. This hypothesis can be verified by future modeling efforts that employ improved numerical algorithms. 

\section{Conclusion and Discussions}\label{sec:discussion}
In this study, we introduce a 2D framework to study the mineral-vapor circulation on rotating lava planets for the first time. Our computations confirm that the rotation of lava planets can generate a significant circulation flow $u$ driven by the Coriolis effect, disrupting the symmetry around the substellar point, in line with the conjectures posited by \citet{Castan2011}. We further investigate the influence of planetary rotation on atmospheric circulations over a wide range of external parameters, and our results are useful for understanding the exchange between the mineral-vapor atmosphere and the surface on lava planets with small nondimenional angular velocity $\tilde{\omega}$. 

In this work, we assume a radiatively transparent atmosphere similar to many early studies \citep[e.g.,][]{Castan2011,Kite2016,nguyen2020k2141b,Kang2021}, so that the temperature profile follows the moist adiabat, which determines the nondimensional parameter $\beta = {R^\ast}/{(R^\ast + \mu C_p)}$ in our momentum equation (Eq.~\ref{eq:2}). According to recent work by \citet{nguyen2022uv}, the absorption of ultraviolet radiation from the host star by the mineral vapor atmosphere can be strong enough to alter the thermal stratification of the atmosphere, resulting in a higher $\beta$ value and faster winds than in a transparent atmosphere. Then the perturbative approach with small parameter expansions we used in this work should work even better as the flow's Rossby number increases. We speculate that inclusion of the real-gas radiative transfer process in our model is not likely to qualitatively alter the behavior of the asymmetric components.

In the future, three-dimensional simulations are needed to fully understand the mineral-vapor circulations on lava planets. The numerical simulations require incorporating the necessary physical schemes in a dynamical core with a similar approach to the study of atmospheric dynamics with a non-dilute condensible component in \citet{pierrehumbert2016nondilute}, and the main challenge lies in the dynamic core's ability to stably handle transonic flows. The three-dimensional model will help to assess our two-dimensional framework presented in this paper, while also facilitating the investigation of various atmospheric processes on lava planets with mineral-vapor atmospheres that cannot be studied in 1D or 2D models, such as convection and associated cloud formation as suggested by \citet{K2}, the influence of minor species as suggested by \citet{Kite2016}, potential formation of shock waves, and temporal variability of the climate.

\begin{acknowledgements}
The authors thank Cheng Li, Lile Wang, and Jun Yang for helpful discussions. F.D. acknowledge funding support from
the Fundamental Research Funds for the Central Universities (Peking University).
\end{acknowledgements}

\software{NumPy (Oliphant, 2006),
    SciPy (Virtanen et al., 2020),
    Matplotlib (Hunter, 2007). }

\appendix

\section{Calculation of the first-order condensation flux $\tilde{\delta D}$}
\label{Appendix:condensation}
When atmospheric condensation occurs in our model, we simply assume that the surface pressure cannot exceed the saturation vapor pressure. Suppose that the condensation zone covers a horizontal domain from the tidally locked latitude $\theta-\delta \theta$ to $\theta+\delta \theta$. Then the condensation flux $D$ must satisfy the saturation condition 
\begin{eqnarray}
    {P}^{(0)} = {P}_{sat} (T^{(0)}), \quad  {P}^{(0)} + \omega  a {P}_1 \sin \phi = {P}_{sat} (T^{(0)}+\omega a T_1 \sin \phi).
\end{eqnarray}
The saturation condition at the first order can be approximately rearranged as
\begin{eqnarray}
    \frac{\text{d} P_1}{\text{d}\theta} \approx \frac{\text{d} P_{sat}}{\text{d} T}\rvert_{T=T^{(0)}} \frac{\text{d} T_1}{\text{d} \theta} + \frac{\text{d}}{\text{d}\theta} \left(\frac{\text{d}P_{sat}}{\text{d}T}\rvert_{T=T^{(0)}}\right) T_1. \label{eq:A:trans2}
\end{eqnarray}
We find that the second term on the right hand side is much smaller than the first term in our simulations, so the second term is ignored.

The condensation fluxes at the zeroth and first orders are determined by the saturation condition and the governing equations at the zeroth and first orders, respectively. The calculation of $D^{(0)}$ has been elucidated in the work of \cite{Kang2021}. Here we will only discuss the condensation flux at the first order.
If the first-order solutions at $\theta$ are known, then we define a vector $\mathbf{e}$ as
\begin{eqnarray}
 \begin{pmatrix}
e_1 \\
e_2 \\
e_3 \\
e_4
\end{pmatrix}
&= \tilde{\Xi_2}^{-1} \left[ \tilde{\Xi_1} \begin{pmatrix}
\tilde{u_1} \\
\tilde{V_1} \\
\tilde{P_1} \\
\tilde{T_1}
\end{pmatrix} + \left( \frac{\text{d} \tilde{\Xi_2}}{\text{d} \theta} \right) \begin{pmatrix}
\tilde{u_1} \\
\tilde{V_1} \\
\tilde{P_1} \\
\tilde{T_1}
\end{pmatrix} + \begin{pmatrix}
2\tilde{V}^{(0)} \tilde{P}^{(0)} \sin^2 \theta \\
0 \\
0 \\
0
\end{pmatrix} \right],
\label{eq:A:dpdt}
\end{eqnarray}
and $\tilde{\delta D}(\theta)$ can be expressed by the elements of vector $\mathbf{e}$ as
\begin{eqnarray}
    \tilde{\delta D} = \frac{e_3 - \left( \frac{\text{d} \tilde{P}_{\text{sat}}}{\text{d} \tilde{T}} \right) e_4}{\beta \tilde{P}^{(0)} + \left( \frac{\text{d} \tilde{P}_{\text{sat}}}{\text{d} \tilde{T}} \right) \left( \beta \tilde{T}^{(0)} - (\tilde{V}^{(0)})^2 \right)} \frac{\tilde{P}^{(0)} \tilde{V}^{(0)}}{\tilde{L}} \left[ (1 - \beta) (\tilde{V}^{(0)})^2 - \beta \tilde{T}^{(0)} \right].
    \label{eq:A:dpdt2}
\end{eqnarray}

Eq.~\ref{eq:A:dpdt2} is clearly valid within the condensation zone. At the boundaries of the condensation zone, corrections have to be made.  
If atmospheric condensation first occurs at the tidally locked latitude $\theta_i$, but $P^{(0)} < P_{\text{sat}} \left( T^{(0)} \right)$ for $\theta_+ > \theta_i$, where $\theta_+$ resides within the positive vicinity of $\theta_i$, then Equ.~\ref{eq:A:dpdt2} still works as long as the transport flow $V$ is positive.
However, if $P^{(0)} < P_{\text{sat}} \left( T^{(0)} \right)$ for $\theta_- < \theta_i$, where $\theta_-$ resides within the negative vicinity of $\theta_i$, then atmospheric condensation will lead to jumps or discontinuities in the first-order solutions, as discussed in Section~\ref{subsec:global}. A representation of the Dirac $\delta$ function needed be introduced to calculate $\tilde{\delta D}$ at this point where condensation first occurs. The jumps of the first-order solution at the condensation boundary should satisfy that 
\begin{eqnarray}
    \tilde{\Xi}_2 \left[
\begin{pmatrix}
\tilde{u_1} \\
\tilde{V_1} \\
\tilde{P_1} \\
\tilde{T_1}
\end{pmatrix}_{\theta_i} -
\begin{pmatrix}
\tilde{u_1} \\
\tilde{V_1} \\
\tilde{P_1} \\
\tilde{T_1}
\end{pmatrix}_{\theta_-}
    \right ]= 
\begin{pmatrix}
0 \\
0 \\
0 \\
\tilde{L} \sin \theta_i \int_{\theta_-}^{\theta_i} \tilde{\delta D'} \text{d} \theta 
\end{pmatrix}.
\label{eq:A:delta}
\end{eqnarray}
Eq.~\ref{eq:A:delta} confirms that the circulation wind $u_1$ is continuous at the condensation boundary, consistent with the green curve in Figure~\ref{fig:show}. The first-order condensation flux at the condensation boundary can be determined by other first-order solutions at $\theta_-$ and the saturation condition at the condensation boundary $\theta_i$
\begin{eqnarray}
\tilde{\delta D'} = \sum_{\theta \rightarrow \theta_{i}^{-}} \Lambda_i \delta(\theta - \theta_i),
\label{eq:A:delta2}
\end{eqnarray}
where $\theta_i$ is the tidally locked latitude of the condensation boundary, and 
\begin{eqnarray}
\Lambda_i = -\frac{ \tilde{P_{1}}(\theta_-) - \left( \frac{\text{d} \tilde{P_{\text{sat}}}}{\text{d} \tilde{T}} \right) \tilde{T_{1}}(\theta_-)}{\beta \tilde{P}^{(0)} + \left( \frac{\text{d} \tilde{P_{\text{sat}}}}{\text{d} \tilde{T}} \right) (\beta \tilde{T}^{(0)} - (\tilde{V}^{(0)})^2)} \frac{\tilde{P}^{(0)} \tilde{V}^{(0)}}{\tilde{L}} \left[ (1 - \beta) (\tilde{V}^{(0)})^2 - \beta \tilde{T}^{(0)} \right]. \label{eq:A:lambda}
\end{eqnarray}



\section{Boundary conditions at the substellar point}
\label{Appendix:boundary condition}
\begin{figure}[htbp]
  \centering
  \begin{minipage}{0.49\textwidth}
    \includegraphics[width=\textwidth]{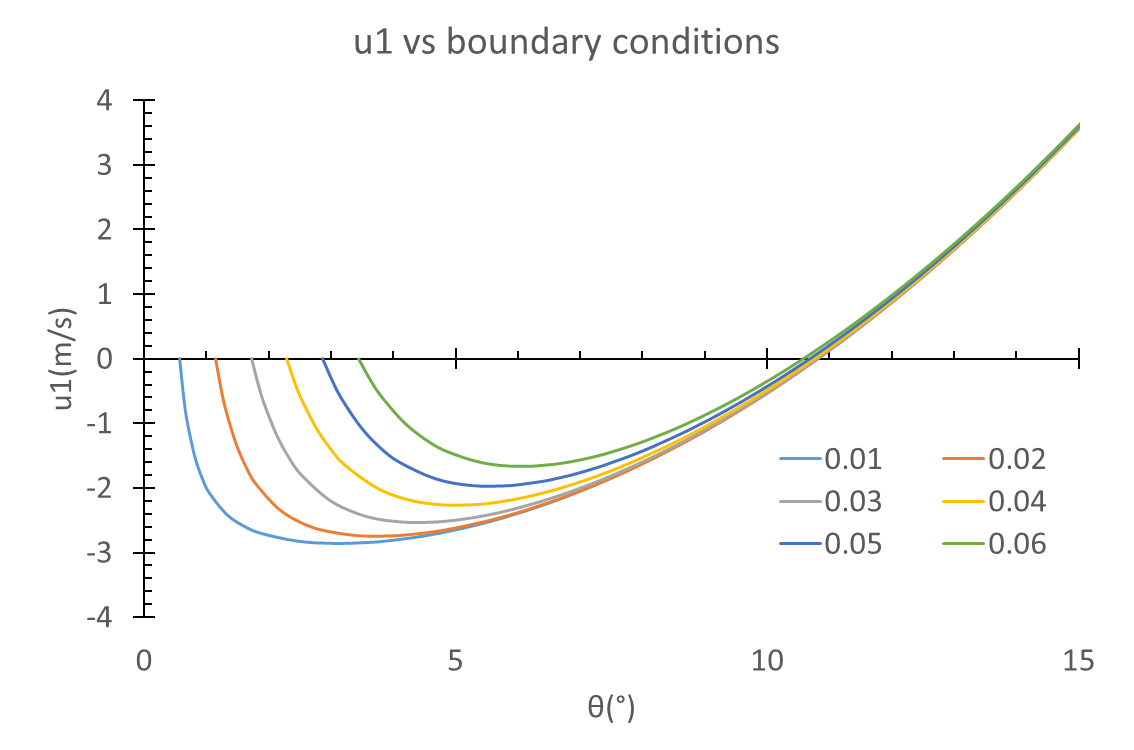}
  \end{minipage}%
  \hfill
  \begin{minipage}{0.49\textwidth}
    \includegraphics[width=\textwidth]{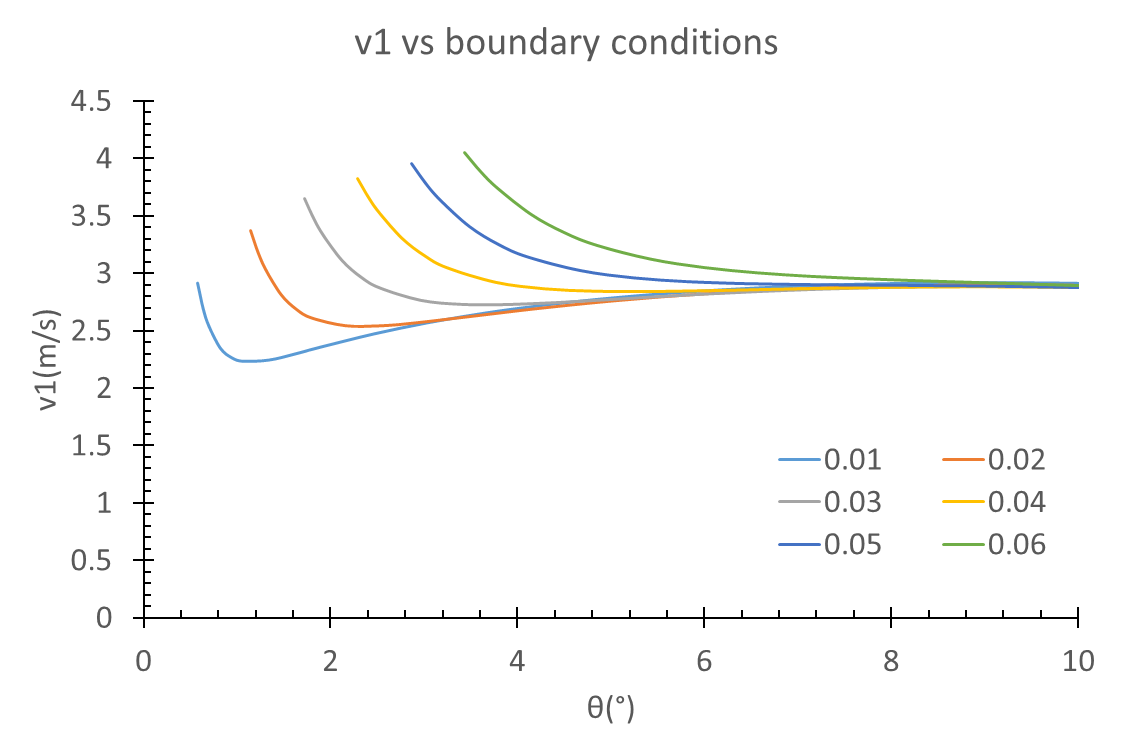}
  \end{minipage}
  
  \medskip 

  \begin{minipage}{0.49\textwidth}
    \includegraphics[width=\textwidth]{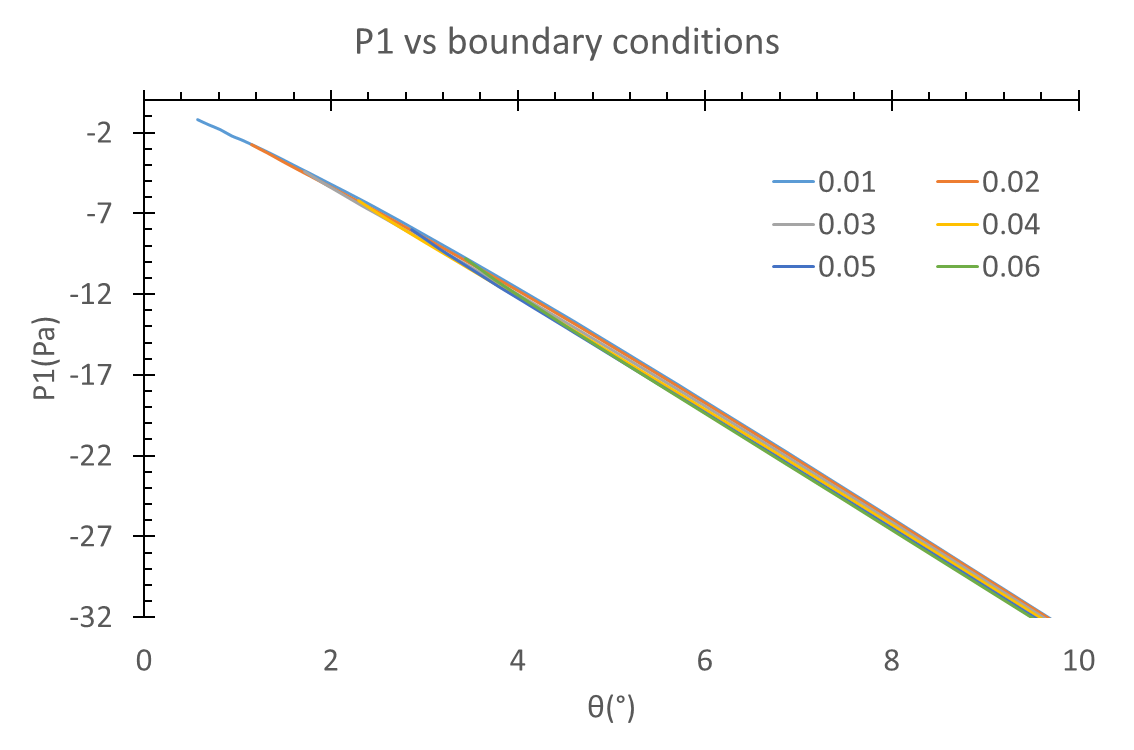}
  \end{minipage}%
  \hfill
  \begin{minipage}{0.49\textwidth}
    \includegraphics[width=\textwidth]{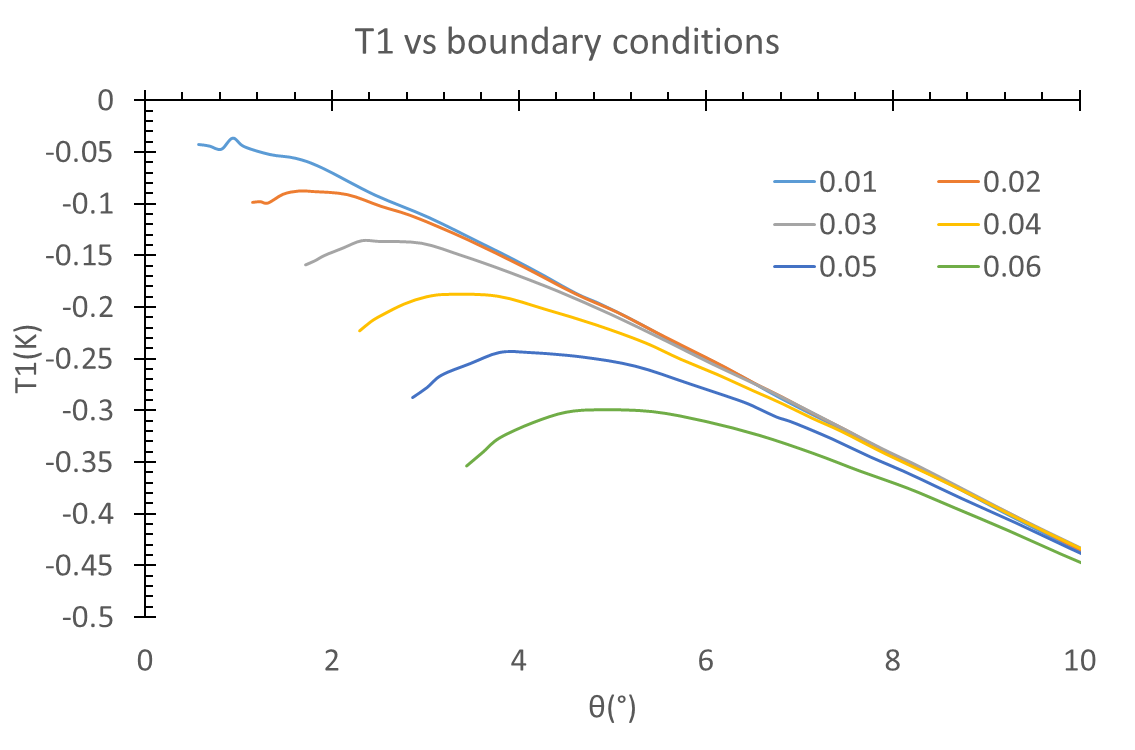}
  \end{minipage}

  \caption{First-order solutions with various starting point $\theta_c$ for numerical integration in our standard simulation. 
  The numerical solutions beyond $\theta>10^\circ$ are insensitive to the choice of $\theta_c$. In each panel, the light blue, red, gray, yellow, dark blue, green color represent $\theta_c = 0.01,0.02,0.03,0.04,0.05,0.06$ radians, respectively.}
   \label{fig:boundary}
\end{figure}

Given that $\tilde{V}^{(0)} =0 $ at $\theta = 0$ \citep{Kang2021}, the first-order governing equations (Eq.~\ref{eq:2d1}) at the substellar point becomes
\begin{eqnarray}
    \frac{\text{d} \mathbf{b}}{\text{d} \theta}\vert_{\theta=0} = -\tilde{\Xi}_1|_{\theta = 0}
    \begin{pmatrix}
    \tilde{u_1} \\\tilde{V_1} \\\tilde{P_1} \\\tilde{T_1}
    \end{pmatrix}_{\theta = 0} = 
        \left(\begin{matrix}
        0 & 0 & -\beta \tilde{T}^{(0)} & -\beta \tilde{P}^{(0)} \\
        -\tilde{P}^{(0)} & 0 & 0 & 0\\
        0 & 0 & \beta \tilde{T}^{(0)} & \beta \tilde{P}^{(0)} \\
        -\tilde{T}^{(0)}\tilde{P}^{(0)} & 0 & 0 & 0
    \end{matrix}\right)_{\theta = 0\circ}
\begin{pmatrix}
    \tilde{u_1} \\\tilde{V_1} \\\tilde{P_1} \\\tilde{T_1}
\end{pmatrix}_{\theta = 0},
\label{eq:B:2}
\end{eqnarray}
where the vector $\mathbf{b}$ is defined as
\begin{eqnarray}
    \mathbf{b} = \tilde{\Xi}_2
    \begin{pmatrix}
        \tilde{u_1} \\\tilde{V_1} \\\tilde{P_1} \\\tilde{T_1}
    \end{pmatrix} .
    \label{eq:B:b}
\end{eqnarray}
When $\theta \rightarrow 0, \tilde{\Xi}_2 \rightarrow 0$, and so is the vector $\mathbf{b}$. 
All boundary condition degenerate at substellar point and the first-order variables cannot be deduced by the vector $\mathbf{b}$ at the substellar point. However, the derivative of $\mathbf{b}$ with respect to $\theta$ at the substellar point can still be expressed as
\begin{eqnarray}
    \frac{\text{d} \mathbf{b}}{\text{d} \theta}|_{\theta =0}  = 
    \left( \begin{matrix}
    0 & 0 & 0 & 0 \\
    0 & \tilde{P}^{(0)} & 0 & 0 \\
    0 & 0 & \beta \tilde{T}^{(0)} & \beta \tilde{P}^{(0)} \\
    0 & \tilde{T}^{(0)}\tilde{P}^{(0)} & 0 & 0
    \end{matrix} \right)_{\theta = 0}
\begin{pmatrix}
   \tilde{u_1} \\\tilde{V_1} \\\tilde{P_1} \\\tilde{T_1}
\end{pmatrix}_{\theta = 0},
\label{eq:B:3}
\end{eqnarray}
Combing Eq.~\ref{eq:B:2} and \ref{eq:B:3} yields 
\begin{eqnarray}
    \frac{\text{d}\mathbf{b}}{\text{d}\theta} =
    \begin{pmatrix}
        0 \\\tilde{P}^{(0)}\rvert_{\theta = 0} \tilde{V}_{1,i} \\0 \\ \tilde{P}^{(0)}\rvert_{\theta = 0} \tilde{T}_{s0} \tilde{V}_{1,i}
    \end{pmatrix},
\end{eqnarray}
where $\tilde{V}_{1,i}$ is the boundary value of $\tilde{V}_1$ at the substellar point $\theta = 0$. When solving the first-order governing equations (Eq.~\ref{eq:2d1}), we choose $\theta = \theta_c \ll 1$ slightly deviating from the substellar point as the starting point for numerical integration, because the coefficient matrix $\tilde{\Xi}_2$ reduces to zero at the substellar point. Then the boundary condition can be written as
\begin{eqnarray}
    \mathbf{b}\rvert_{\theta = \theta_c} =
    \begin{pmatrix}
        0 \\\tilde{P}^{(0)}\rvert_{\theta = 0} \tilde{V}_{1,i}  \theta_c \\0 \\ \tilde{P}^{(0)}\rvert_{\theta = 0} \tilde{T}_{s0} \tilde{V}_{1,i} \theta_c
    \end{pmatrix}
\end{eqnarray}
We have tried six different values for $\theta_c$ for numerical integration, and adjusted the value of $\tilde{V}_{1,i}$ correspondingly to ensure that the steady-state solutions pass through the sonic point smoothly. Figure~\ref{fig:boundary} shows that the first-order solutions beyond $\theta > 10^\circ$ are insensitive to the choice of $\theta_c$, because of the sonic point constraint.

\section{Residual terms of perturbation solutions}
\label{Appendix:system error}

\begin{figure}
\centering
\includegraphics[width=1\linewidth]{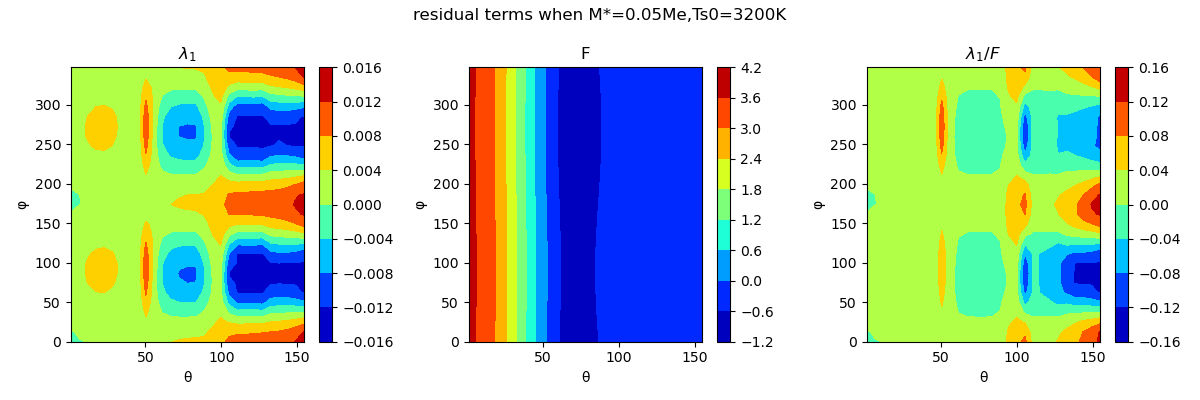}
\caption{Surface distribution of the residual term $\lambda_1$ in Eq.~\ref{eq:residual1} (left panel), mass exchange flux between the surface and atmosphere $F$ (middle panel), and $\lambda_1/F$ (right panel) in our standard simulation. $\lambda_1$ is mainly made of wave-number-two component along the tidally locked longitude $phi$ and is much smaller than $F$ when $F$ is not close to 0.} \label{fig:residual1}
\end{figure}

We ignore the higher-order terms when solving the basic equations in this work. Their contribution can be assessed by comparison with the zeroth- and first-order solutions. These comparisons help verify the appropriateness of our simplification.  We define the residual terms as follows when substituting the zeroth- and first-order solutions (Eq.~\ref{eq:define4}) into the original basic equations (Eqs.~\ref{eq:1}-\ref{eq:3})
\begin{eqnarray}
& \nabla \cdot \left( \frac{\mathbf{V} P}{g} \right) - F &= \lambda_1 \label{eq:residual1},\\
& \nabla \cdot \left( \frac{\mathbf{V} \mathbf{V} {P}}{g} \right) +  \nabla \left( \frac{\beta C_p T P}{g} \right) - {F}_- \mathbf{V} +2\omega \sin \theta \cos \phi (\mathbf{n}\times \mathbf{V}) \frac{P}{g} &= \mathbf{\lambda_2} \label{eq:residual2},\\
& \nabla \cdot \left[ \left( \frac{|\mathbf{V}|^2}{2} + C_p T \right) \frac{\mathbf{V} {P}}{g} \right] - D {L} - {F}_- \left( \frac{|\mathbf{V}|^2}{2} + C_p T \right) - {F}_+ C_p T_s &= \lambda_3 \label{eq:residual3}.
\end{eqnarray}
Then the residual term $\lambda_1$ in Eq.~\ref{eq:residual1} can be expressed as
\begin{eqnarray}
    \lambda_1 = \frac{\text{d}}{a \sin{\theta} \text{d} \theta} \left(\frac{V^{(1)}P^{(1)} \sin{\theta}}{g}\right) + \frac{\text{d}}{a \sin{\theta} \text{d} \phi} \left(\frac{u^{(1)}P^{(1)}}{g}\right) - \left[F(T^{(0)}+T^{(1)},P^{(0)}+P^{(1)}) - F^{(0)} - F_{,P}P^{(1)} - F_{,T}T^{(1)}\right]
\end{eqnarray}
where $u^{(1)} = -\omega a u_1 \cos \phi, V^{(1)} = \omega a V_1 \sin \phi,
    P^{(1)} = \omega a P_1  \sin \phi,
    T^{(1)} = \omega a T_1  \sin \phi$.
The surface distribution of $\lambda_1$ exhibits a wave-number-two feature along the tidally locked longitude $\phi$, as shown in Figure~\ref{fig:residual1}. The magnitude of $\lambda_1$ is much smaller than the flux of mass exchange between the surface and the atmosphere $F$ when $\theta \leq 150^\circ$.

\begin{figure}
\centering
\includegraphics[width=1\linewidth]{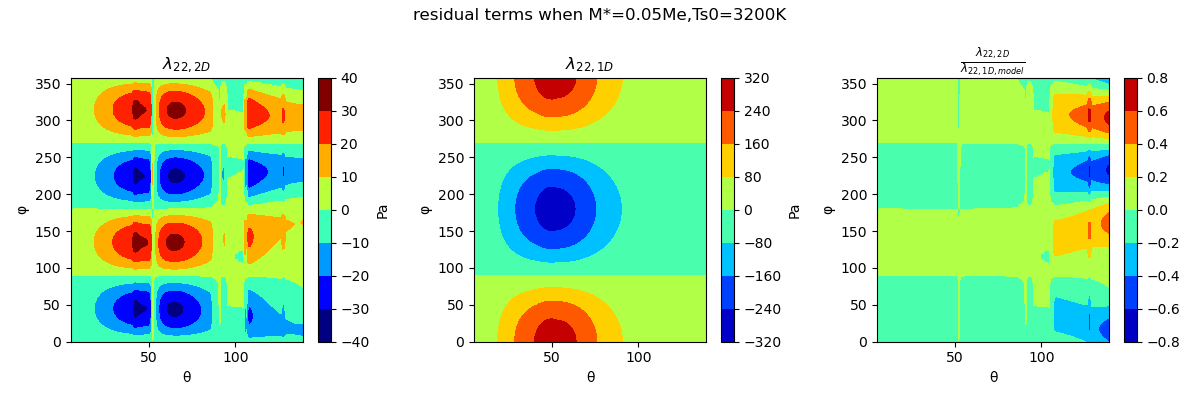}
\caption{ Surface distribution of the residual term $\lambda_{22,2D}$ in Eq.~\ref{eq:residual2} (left panel), Coriolis term in the first-order momentum equation $\lambda_{22,1D}$ (middle panel), and their ratio $\lambda_{22,2D}/\lambda_{22,1D}$ (right panel) in our standard simulation. $\lambda_{22,2D}$ is mainly made of wave-number-two component along the tidally locked longitude $\phi$ and is much smaller than $\lambda_{22,1D}$ when $\theta \leq 110^{\circ}$.}
\label{fig:residual}
\end{figure}

The residual term in the momentum equation can be decomposed as $\mathbf{\lambda_2} = \lambda_{21,2D}\mathbf{e_\theta} + \lambda_{22,2D}\mathbf{e_\phi}$, and the component along the tidally locked longitude is
\begin{eqnarray}
\begin{array}{ll}
    \lambda_{22,2D} = & \frac{\text{d}}{a \sin{\theta} \text{d} \theta} \left[\frac{u^{(1)} ( V^{(1)}P^{(0)} + P^{(1)}V^{(0)} + V^{(1)}P^{(1)} )}{g} \sin{\theta}\right] + \frac{\text{d}}{a \sin{\theta} \text{d}\phi} \left[\frac{(u^{(1)})^2 P^{(0)}+\beta C_p T^{(1)}P^{(1)}}{g}  \right] \\
    & - \left[F_-(T^{(0)}+T^{(1)},P^{(0)}+P^{(1)}) - F_-^{(0)}\right]u^{(1)} + \frac{2\omega (P^{(0)} V^{(1)}+ P^{(1)} V^{(0)})\sin \theta \cos\phi}{g} .
\end{array}    \label{eq:lambda222D}
\end{eqnarray}
The Coriolis term in the first-order momentum equation is 
\begin{equation}
    \lambda_{22,1D} =  \frac{2\omega P^{(0)} V^{(0)} \sin \theta \cos\phi}{g} 
    \label{eq:lambda221D}    
\end{equation}
Similar to the residual term in the mass equation (Eq.~\ref{eq:residual1}), the surface distribution of $\lambda_{22,2D}$ is also dominated by the wave-number-two component along the tidally locked longitude $\phi$, as shown in Figure~\ref{fig:residual}. The magnitude of $\lambda_{22,2D}$ is much smaller than the Coriolis term $\lambda_{22,1D}$ when $\theta \leq 110^{\circ}$. But when $\theta \geq 130^{\circ}$, $\lambda_{22,2D}$ is comparable to $\lambda_{22,1D}$, indicating that higher-order terms are required to describe the asymmetric flow pattern there. 

\section{Linear stability to small perturbations in the first-order solutions}
\label{Appendix:relaxation}
\begin{figure}[htbp]
  \centering
  \begin{minipage}{0.49\textwidth}
    \includegraphics[width=\textwidth]{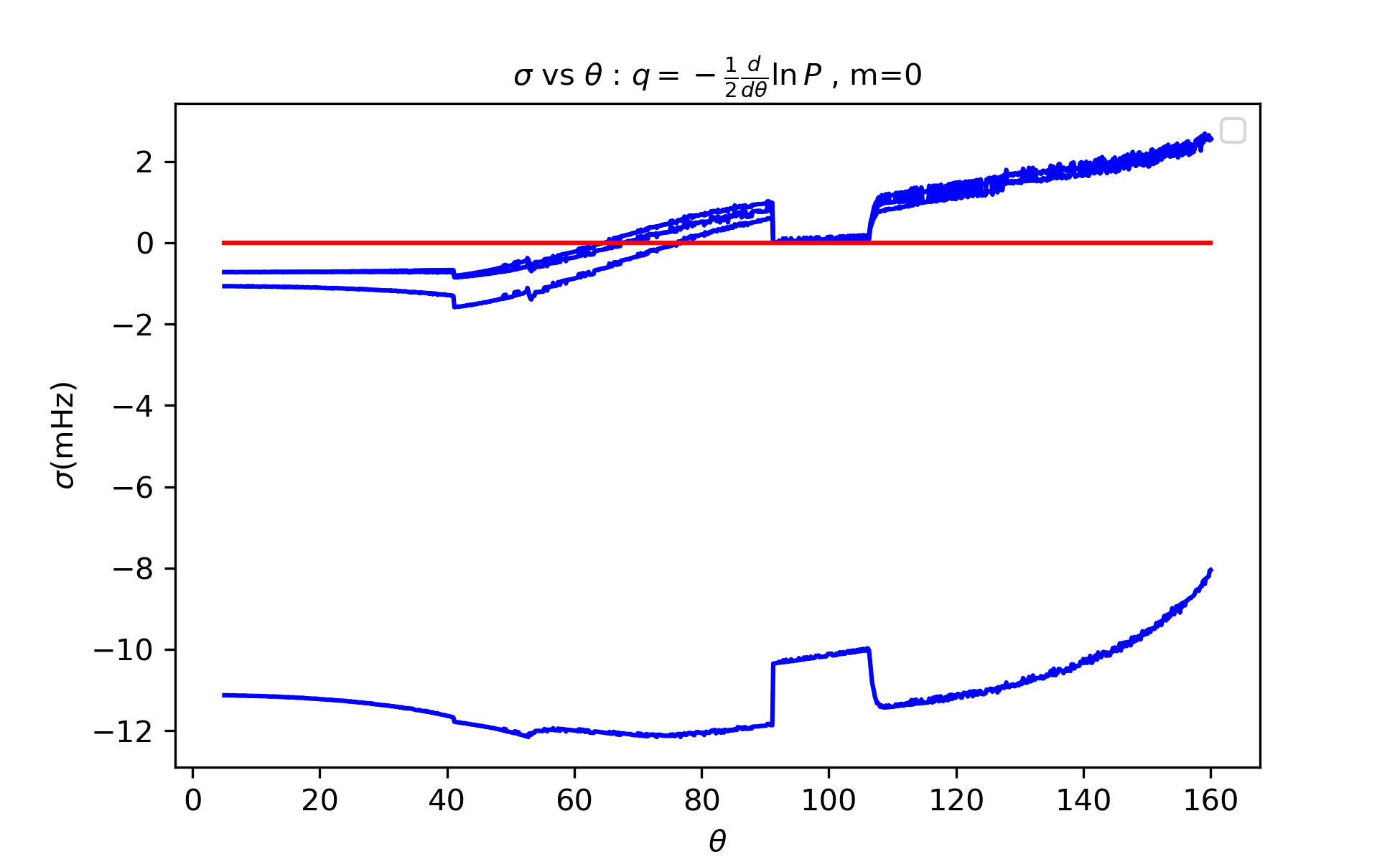}
  \end{minipage}%
  \hfill
  \begin{minipage}{0.49\textwidth}
    \includegraphics[width=\textwidth]{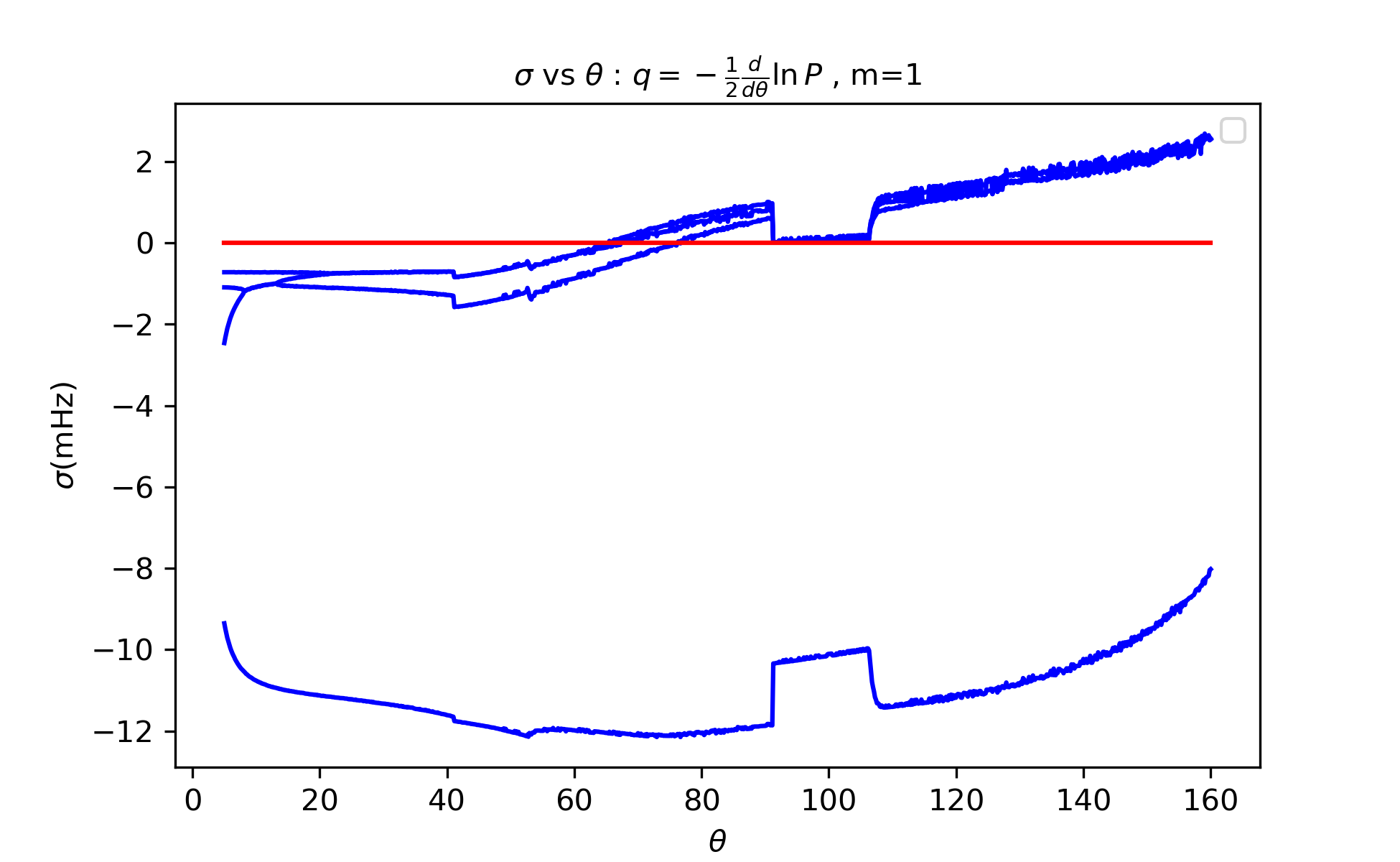}
  \end{minipage}
  

  
  \caption{Real part of the complex frequency $\sigma$ as a function of the tidally locked latitude $\theta$ in our standard simulation, when $m = 0$ (left panel) and $m=1$ (right panel). The four blue lines in each panel represent the four solutions of $\sigma$ in Eq.~\ref{eq:dispersion}. The zero value of $\sigma$ is marked by the red horizontal lines. }
  \label{fig:relax} 
\end{figure}

In this paper, we focus on obtaining steady-state solutions, consistent with previous studies. Here, we add small perturbations to the first-order solutions, examine their linear stability property in a crude approach, and check whether the climate equilibrium state we find is stable or not. First, the basic equations that varies with time are
\begin{align}
& \nabla \cdot \left( \frac{\mathbf{V} P}{g} \right) - F &=& - \frac{\partial}{\partial t}\left(\frac{P}{g} \right) \label{eq:relax1}, \\
& \nabla \cdot \left( \frac{\mathbf{V} \mathbf{V} {P}}{g} \right) +  \nabla \left( \frac{\beta C_p T P}{g} \right) - {F}_- \mathbf{V} +2\omega \sin \theta \cos \phi (\mathbf{n}\times \mathbf{V}) \frac{P}{g} &=& - \frac{\partial}{\partial t}\left( \frac{\mathbf{V}P}{g} \right)
\label{eq:relax2}, \\
& \nabla \cdot \left[ \left( \frac{|\mathbf{V}|^2}{2} + C_p T \right) \frac{\mathbf{V} {P}}{g} \right] - D {L} - {F}_- \left( \frac{|\mathbf{V}|^2}{2} + C_p T \right) - {F}_+ C_p T_s &=& - \frac{\partial}{\partial t}\frac{({|\mathbf{V}|^2}/{2} + C_p T)P}{g} \label{eq:relax3}.
\end{align}
Then we prescribe nondimensional perturbations that oscillate around the tidally locked longitude $\phi$ with wavenumber of $m$
\begin{align}
\begin{pmatrix}
\delta u(\theta, \phi, t)\\
\delta V(\theta, \phi, t)\\
\delta P(\theta, \phi, t)\\
\delta T(\theta, \phi, t)
\end{pmatrix} = 
\begin{pmatrix}
- \delta u' \cos (m\phi)\\
\delta V' \sin (m\phi)\\
\delta P'\sin (m\phi)\\
\delta T'\sin (m\phi)
\end{pmatrix} = 
\begin{pmatrix}
- \tilde{u}_0 \cos({m\phi}) \\ 
\tilde{V}_0 \sin({m\phi}) \\ 
\tilde{P}_0 \sin({m\phi}) \\ 
\tilde{T}_0 \sin({m\phi}) 
\end{pmatrix} 
\exp[q(\theta)\theta+\sigma t],
\label{eq:relax_define_1}
\end{align}
where 
\begin{align}
\begin{pmatrix}
\delta u' \\
\delta V' \\
\delta P'\\
\delta T'
\end{pmatrix} = 
\begin{pmatrix}
 \tilde{u}_0  \\ 
\tilde{V}_0  \\ 
\tilde{P}_0  \\ 
\tilde{T}_0 
\end{pmatrix} 
\exp[q(\theta)\theta+\sigma t],
\label{eq:relax_define_2}
\end{align}
and the real part of the complex frequency $\sigma$ represents exponential growth (positive) or decay (negative) of the initial perturbation.
The dispersion relation can be obtained by substituting the perturbations into the time-varying nondimensional equations (Eq.~\ref{eq:relax1}-\ref{eq:relax3})
\begin{align}
\tilde{\Xi}_{1m} 
\begin{pmatrix}
\delta u' \\
\delta V' \\
\delta P' \\
\delta T'
\end{pmatrix}
+ \frac{\partial}{\partial \theta} \left[ \tilde{\Xi}_{2m} \begin{pmatrix}
\delta u' \\
\delta V' \\
\delta P' \\
\delta T'
\end{pmatrix}
\right] &= - \frac{\partial}{\partial t} \left[ {\sin \theta}
\begin{pmatrix}
-P^{(0)}\delta u' \\
\delta P' \\
\tilde{P}^{(0)}\delta V' + \tilde{V}^{(0)}\delta P' \\
({(\tilde{V}^{(0)})^2}/{2} +  \tilde{T}^{(0)}) \delta P' + (\tilde{V}^{(0)} \delta V' +  \delta T') P^{(0)} 
\end{pmatrix}\right] = -\sigma \tilde{\Xi}_{3m} \begin{pmatrix}
\delta u' \\
\delta V' \\
\delta P' \\
\delta T'
\end{pmatrix}
\label{eq:relaxiation},
\end{align}
where the coefficient matrices are
\begin{align}
    & \tilde{\Xi}_{1m} & =& \begin{pmatrix}
  \tilde{F}_-^{(0)} \sin \theta & 0 & m \beta \tilde{T}^{(0)} & m \beta \tilde{P}^{(0)} \\
  m\tilde{P}^{(0)} & 0 & -\tilde{F}_{,\tilde{P}} \sin\theta & -\tilde{F}_{,\tilde{T}} \sin\theta \\
  m\tilde{V}^{(0)} \tilde{P}^{(0)} & -\tilde{F}_{-}^{(0)} \sin \theta & -\beta \tilde{T}^{(0)} \cos \theta -\tilde{F}_{-,\tilde{P}} \tilde{V}^{(0)} \sin \theta & -\beta \tilde{P}^{(0)} \cos \theta -\tilde{F}_{-,\tilde{T}} \tilde{V}^{(0)} \sin \theta \\
  m[\frac{(\tilde{V}^{(0)})^2}{2} + \tilde{T}^{(0)}] \tilde{P}^{(0)} & -\tilde{F}_-^{(0)} \tilde{V}^{(0)} \sin \theta & 
  \begin{array}{c}
  \substack{
  [- \tilde{F}_{+,\tilde{P}} \tilde{T}_s - \\   \tilde{F}_{-,\tilde{P}} (\tilde{T}^{(0)} + \frac{(\tilde{V}^{(0)})^2}{2})]\sin \theta}
  \end{array} & 
  \begin{array}{c}
  \substack{[- \tilde{F}_-^{(0)} - \tilde{F}_{+,\tilde{T}} \tilde{T}_s  - \\ \tilde{F}_{-,\tilde{T}} (  \tilde{T}^{(0)} + \frac{(\tilde{V}^{(0)})^2}{2})] \sin \theta}
  \end{array}
\end{pmatrix}, 
  \label{eq:xi1m} \\
  &  \tilde{\Xi}_{2m} & =&\begin{pmatrix}
  -\tilde{V}^{(0)} \tilde{P}^{(0)} & 0 & 0 & 0 \\
  0 & \tilde{P}^{(0)} & \tilde{V}^{(0)} & 0 \\
  0 & 2\tilde{V}^{(0)} \tilde{P}^{(0)} & (\tilde{V}^{(0)})^2 + \beta \tilde{T}^{(0)} & \beta \tilde{P}^{(0)} \\
  0 & (3(\tilde{V}^{(0)})^2/2 + \tilde{T}^{(0)}) \tilde{P}^{(0)} & ((\tilde{V}^{(0)})^2/2 + \tilde{T}^{(0)}) \tilde{V}^{(0)} & \tilde{V}^{(0)} \tilde{P}^{(0)}
\end{pmatrix} \sin\theta, \label{eq:xi2m} \\
    &\tilde{\Xi}_{3m} & =& \left( \begin{matrix}
    -\tilde{P}^{(0)} & 0 & 0 & 0 \\
    0 & 0 & 1 & 0 \\
    0 & \tilde{P}^{(0)} & \tilde{V}^{(0)} & 0 \\
    0 & \tilde{V}^{(0)}\tilde{P}^{(0)} & (\tilde{V}^{(0)})^2 + \tilde{T}^{(0)} & \tilde{P}^{(0)}
\end{matrix} \right) \sin \theta.
\label{eq:xi3m}
\end{align}
The existence of nontrivial solution requires that 
\begin{align}
    \det \tilde{\Xi}_t(q, \sigma) = 0, \quad \text{where}\  \tilde{\Xi}_t = \tilde{\Xi}_{1m} + q(\theta) \tilde{\Xi}_{2m}  + \frac{\text{d} \tilde{\Xi}_{2m}}{\text{d}\theta} + \sigma \tilde{\Xi}_{3m}. 
    \label{eq:dispersion}
\end{align}
We further assume that the amplitude of perturbations varies with the background surface pressure, and follows that $\tilde{P}^{(0)} \exp[2q(\theta)\theta] \sim const.$, which has also been used for theoretical studies of vertically propagating gravity waves on Earth \citep[e.g.,][]{lindzen1967thermaltide,lindzen1981gravitywave}. Then $q$ can be roughly estimated as
\begin{align}
    q(\theta) \sim -\frac{1}{2}\frac{\text{d}}{\text{d}\theta} \ln{\tilde{P}^{(0)}(\theta)}.
\end{align}
Once $q(\theta)$ is known, the complex frequency $\sigma(\theta)$ can be solved by the dispersion relation (Eq.~\ref{eq:dispersion}). Figure~\ref{fig:relax} shows the real part of the complex frequency $\sigma$ as a function of the tidally locked latitude $\theta$ for $m=0,1$ in our standard simulation. Eq.~\ref{eq:dispersion} is quadratic, and the real part of the four solutions stays negative when $\theta \leq 80^\circ$. It indicates that the first-order solution in steady state we found in this work is stable there and is consistent with our analysis on the residual terms in Appendix~\ref{Appendix:system error}. 

The instability on the nightside does not necessarily imply that steady flow is improbable on the night hemisphere, and it could be suppressed if higher-order terms are included in the 2D model. Moreover, even if the flow is unsteady on the night hemisphere, perturbations or information cannot be transported upstream to the day hemisphere because the supersonic transport velocity $V$ at the terminator is much higher than the speed of sound \citep{Kang2021}.


\bibliography{sample631}{}

\begin{thebibliography}{}
\expandafter\ifx\csname natexlab\endcsname\relax\def\natexlab#1{#1}\fi
\providecommand{\url}[1]{\href{#1}{#1}}
\providecommand{\dodoi}[1]{doi:~\href{http://doi.org/#1}{\nolinkurl{#1}}}
\providecommand{\doeprint}[1]{\href{http://ascl.net/#1}{\nolinkurl{http://ascl.net/#1}}}
\providecommand{\doarXiv}[1]{\href{https://arxiv.org/abs/#1}{\nolinkurl{https://arxiv.org/abs/#1}}}

\bibitem[{{Castan} \& {Menou}(2011)}]{Castan2011}
{Castan}, T., \& {Menou}, K. 2011, \apjl, 743, L36, \dodoi{10.1088/2041-8205/743/2/L36}

\bibitem[{{Demory} {et~al.}(2016){Demory}, {Gillon}, {de Wit}, {Madhusudhan}, {Bolmont}, {Heng}, {Kataria}, {Lewis}, {Hu}, {Krick}, {Stamenkovi{\'c}}, {Benneke}, {Kane}, \& {Queloz}}]{Demory2016}
{Demory}, B.-O., {Gillon}, M., {de Wit}, J., {et~al.} 2016, \nat, 532, 207, \dodoi{10.1038/nature17169}

\bibitem[{{Ding} \& {Pierrehumbert}(2018)}]{Ding2018}
{Ding}, F., \& {Pierrehumbert}, R.~T. 2018, \apj, 867, 54, \dodoi{10.3847/1538-4357/aae38c}

\bibitem[{{Guo}(2024)}]{Guo2024}
{Guo}, J.~H. 2024, Nature Astronomy, \dodoi{10.1038/s41550-024-02269-w}

\bibitem[{{Hashimoto}(1990)}]{Hashimoto1990}
{Hashimoto}, A. 1990, \nat, 347, 53, \dodoi{10.1038/347053a0}

\bibitem[{{Herath} {et~al.}(2024){Herath}, {Boukar{\'e}}, \& {Cowan}}]{herath2024thermallava}
{Herath}, M., {Boukar{\'e}}, C.-{\'E}., \& {Cowan}, N.~B. 2024, \mnras, 535, 2404, \dodoi{10.1093/mnras/stae2431}

\bibitem[{{Hu} {et~al.}(2024){Hu}, {Bello-Arufe}, {Zhang}, {Paragas}, {Zilinskas}, {van Buchem}, {Bess}, {Patel}, {Ito}, {Damiano}, {Scheucher}, {Oza}, {Knutson}, {Miguel}, {Dragomir}, {Brandeker}, \& {Demory}}]{Hu2024}
{Hu}, R., {Bello-Arufe}, A., {Zhang}, M., {et~al.} 2024, \nat, 630, 609, \dodoi{10.1038/s41586-024-07432-x}

\bibitem[{{Ingersoll} {et~al.}(1985){Ingersoll}, {Summers}, \& {Schlipf}}]{Ingersoll1985}
{Ingersoll}, A.~P., {Summers}, M.~E., \& {Schlipf}, S.~G. 1985, \icarus, 64, 375, \dodoi{10.1016/0019-1035(85)90062-4}

\bibitem[{{Kang} {et~al.}(2021){Kang}, {Ding}, {Wordsworth}, \& {Seager}}]{Kang2021}
{Kang}, W., {Ding}, F., {Wordsworth}, R., \& {Seager}, S. 2021, \apj, 906, 67, \dodoi{10.3847/1538-4357/abcaa7}

\bibitem[{{Kang} {et~al.}(2023){Kang}, {Nimmo}, \& {Ding}}]{Kang2023}
{Kang}, W., {Nimmo}, F., \& {Ding}, F. 2023, \apjl, 949, L20, \dodoi{10.3847/2041-8213/acd691}

\bibitem[{{Kite} \& {Barnett}(2020)}]{kite2020secondary}
{Kite}, E.~S., \& {Barnett}, M.~N. 2020, Proceedings of the National Academy of Science, 117, 18264, \dodoi{10.1073/pnas.2006177117}

\bibitem[{{Kite} {et~al.}(2016){Kite}, {Fegley}, {Schaefer}, \& {Gaidos}}]{Kite2016}
{Kite}, E.~S., {Fegley}, Bruce, J., {Schaefer}, L., \& {Gaidos}, E. 2016, \apj, 828, 80, \dodoi{10.3847/0004-637X/828/2/80}

\bibitem[{{Koll} \& {Abbot}(2015)}]{koll2015phasecurve}
{Koll}, D. D.~B., \& {Abbot}, D.~S. 2015, \apj, 802, 21, \dodoi{10.1088/0004-637X/802/1/21}

\bibitem[{{Lai} {et~al.}(2024){Lai}, {Yang}, \& {Kang}}]{lai2024lavaocean}
{Lai}, Y., {Yang}, J., \& {Kang}, W. 2024, \psj, 5, 204, \dodoi{10.3847/PSJ/ad7111}

\bibitem[{{Landau} \& {Lifshitz}(1959)}]{landau1959fluidbook}
{Landau}, L.~D., \& {Lifshitz}, E.~M. 1959, {Fluid mechanics}

\bibitem[{{Lindzen}(1967)}]{lindzen1967thermaltide}
{Lindzen}, R.~S. 1967, Quarterly Journal of the Royal Meteorological Society, 93, 18, \dodoi{10.1002/qj.49709339503}

\bibitem[{{Lindzen}(1981)}]{lindzen1981gravitywave}
---. 1981, \jgr, 86, 9707, \dodoi{10.1029/JC086iC10p09707}

\bibitem[{{Merlis} \& {Schneider}(2010)}]{merlis2010tidallylocked}
{Merlis}, T.~M., \& {Schneider}, T. 2010, Journal of Advances in Modeling Earth Systems, 2, 13, \dodoi{10.3894/JAMES.2010.2.13}

\bibitem[{{Nguyen} {et~al.}(2020){Nguyen}, {Cowan}, {Banerjee}, \& {Moores}}]{nguyen2020k2141b}
{Nguyen}, T.~G., {Cowan}, N.~B., {Banerjee}, A., \& {Moores}, J.~E. 2020, \mnras, 499, 4605, \dodoi{10.1093/mnras/staa2487}

\bibitem[{{Nguyen} {et~al.}(2022){Nguyen}, {Cowan}, {Pierrehumbert}, {Lupu}, \& {Moores}}]{nguyen2022uv}
{Nguyen}, T.~G., {Cowan}, N.~B., {Pierrehumbert}, R.~T., {Lupu}, R.~E., \& {Moores}, J.~E. 2022, \mnras, 513, 6125, \dodoi{10.1093/mnras/stac1331}

\bibitem[{{Parker}(1965)}]{parker1965solarwind}
{Parker}, E.~N. 1965, \ssr, 4, 666, \dodoi{10.1007/BF00216273}

\bibitem[{Petzold(1983)}]{petzold1983lsoda}
Petzold, L. 1983, SIAM Journal on Scientific and Statistical Computing, 4, 136, \dodoi{10.1137/0904010}

\bibitem[{{Pierrehumbert} \& {Ding}(2016)}]{pierrehumbert2016nondilute}
{Pierrehumbert}, R.~T., \& {Ding}, F. 2016, Proceedings of the Royal Society of London Series A, 472, 20160107, \dodoi{10.1098/rspa.2016.0107}

\bibitem[{{Pierrehumbert} \& {Hammond}(2019)}]{pierrehumbert2019tidelocked}
{Pierrehumbert}, R.~T., \& {Hammond}, M. 2019, Annual Review of Fluid Mechanics, 51, 275, \dodoi{10.1146/annurev-fluid-010518-040516}

\bibitem[{{Samuel} {et~al.}(2014){Samuel}, {Leconte}, {Rouan}, {Forget}, {L{\'e}ger}, \& {Schneider}}]{Samuel2014observationCoRot7b}
{Samuel}, B., {Leconte}, J., {Rouan}, D., {et~al.} 2014, \aap, 563, A103, \dodoi{10.1051/0004-6361/201321039}

\bibitem[{{Schaefer} \& {Fegley}(2009)}]{SodiumIsTheMost2009}
{Schaefer}, L., \& {Fegley}, B. 2009, \apjl, 703, L113, \dodoi{10.1088/0004-637X/703/2/L113}

\bibitem[{{Schaefer} {et~al.}(2012){Schaefer}, {Lodders}, \& {Fegley}}]{schaefer2012vaporization}
{Schaefer}, L., {Lodders}, K., \& {Fegley}, B. 2012, \apj, 755, 41, \dodoi{10.1088/0004-637X/755/1/41}

\bibitem[{{Showman} \& {Polvani}(2011)}]{showman2011superrotation}
{Showman}, A.~P., \& {Polvani}, L.~M. 2011, \apj, 738, 71, \dodoi{10.1088/0004-637X/738/1/71}

\bibitem[{{Showman} {et~al.}(2012){Showman}, {Wordsworth}, \& {Merlis}}]{showman2012circulation}
{Showman}, A.~P., {Wordsworth}, R.~D., \& {Merlis}, T.~M. 2012, in LPI Contributions, Vol. 1675, Comparative Climatology of Terrestrial Planets, ed. {LPI Editorial Board}, 8090

\bibitem[{Valencia {et~al.}(2010)Valencia, Ikoma, Guillot, \& Nettelmann}]{Valencia2010fateofCoRoT7b}
Valencia, D., Ikoma, M., Guillot, T., \& Nettelmann, N. 2010, Astronomy \& Astrophysics, 516, A20

\bibitem[{{Wordsworth} \& {Kreidberg}(2022)}]{Summary2022}
{Wordsworth}, R., \& {Kreidberg}, L. 2022, \araa, 60, 159, \dodoi{10.1146/annurev-astro-052920-125632}

\bibitem[{{Wordsworth} {et~al.}(2018){Wordsworth}, {Schaefer}, \& {Fischer}}]{wordsworth2018redox}
{Wordsworth}, R.~D., {Schaefer}, L.~K., \& {Fischer}, R.~A. 2018, \aj, 155, 195, \dodoi{10.3847/1538-3881/aab608}

\bibitem[{{Zieba} {et~al.}(2022){Zieba}, {Zilinskas}, {Kreidberg}, {Nguyen}, {Miguel}, {Cowan}, {Pierrehumbert}, {Carone}, {Dang}, {Hammond}, {Louden}, {Lupu}, {Malavolta}, \& {Stevenson}}]{K2}
{Zieba}, S., {Zilinskas}, M., {Kreidberg}, L., {et~al.} 2022, \aap, 664, A79, \dodoi{10.1051/0004-6361/202142912}

\end{thebibliography}
\bibliographystyle{aasjournal}



\end{CJK*}
\end{document}